\documentclass[aps,prb,fleqn,12pt]{revtex4}
\usepackage{epsfig,color}
\usepackage{graphicx}
\usepackage{pdfpages}
\usepackage{braket}
\usepackage{amssymb,amsmath}
\usepackage{hyperref}
\begin{document}
\title{Role of orbital off-diagonal spin and charge condensates in a three orbital model 
for {\boldmath $\rm Ca_2RuO_4$} --- Coulomb renormalized spin-orbit coupling, orbital moment, and tunable magnetic order}
\author{Shubhajyoti Mohapatra, Ritajit Kundu, Ashutosh Dubey, \\ Debasis Dutta, and Avinash Singh}
\email{avinas@iitk.ac.in}
\affiliation{Department of Physics, Indian Institute of Technology, Kanpur - 208016, India}
\date{\today} 
\begin{abstract}
Strongly anisotropic spin-orbit coupling (SOC) renormalization and strongly enhanced orbital magnetic moments are obtained in the fully self consistent approach including the orbital off-diagonal spin and charge condensates. For moderate tetragonal distortion as in $\rm Ca_2 RuO_4$, dominantly planar antiferromagnetic (AFM) order with small canting of moments in and about the crystal $c$ axis are obtained. For reduced tetragonal distortion, we find a tunable regime wherein the magnetic order can be tuned (AFM or FM) by the bare SOC strength and octahedral tilting magnitude. In this regime, with decreasing tetragonal distortion, AFM order is maintained by progressively decreasing octahedral tilting, as observed in $\rm Ca_{2-x}Sr_x RuO_4$. For purely planar order, the only self consistent solution is FM order along crystal $b$ axis, which is relevant for the bilayer ruthenate compound  $\rm Ca_3 Ru_2 O_7$. 
\end{abstract}
\maketitle
\newpage

\section{Introduction}
The ruthenium-based quasi-two-dimensional ($2D$) square-lattice compounds $\rm A_2RuO_4$ (A=Sr,Ca) with $4d^4$ electronic configuration have attracted renewed interest due to the sensitivity of the low-energy physics of these systems to the complex interplay between spin-orbit coupling (SOC), Coulomb interaction terms, tetragonal distortion, and octahedral tilting and rotation. This complex interplay has a crucial role in the gradual transition from strongly correlated metallicity in $\rm Sr_2RuO_4$ to unusual magnetism in $\rm Ca_2RuO_4$,\cite{fatuzzo_PRB_2015,zhang_PRL_2016,jain_NATPHY_2017,souliou_PRL_2017,zhang_PRB_2017,porter_PRB_2018,das_PRX_2018,dietl_APL_2018,kim_PRL_2018,feldmaier_arxiv_2019,gretarsson_PRB_2019,zhang_PRB_2020}
which exhibits coupled spin-orbital excitations with energy decreasing from zone center to zone boundary, and pressure and chemical substitution induced magnetic reorientation transition from antiferromagnetic (AFM) to ferromagnetic (FM) order, which is driven by octahedral de-flattening and accompanied with decreasing octahedral tilting. Recent investigations on related compounds include comparative study of magnetic excitations in $\rm Ca_2RuO_4$ and the bilayer ruthenate $\rm Ca_3Ru_2O_7$ using Resonant Inelastic X-ray Scattering (RIXS),\cite{arx_arxiv_2020} and the co-existence of superconductivity and ferromagnetism in a $\rm Ca_2RuO_4$ nanofilm crystal.\cite{nobukane_SREP_2020}

The ground state in the isoelectronic series $\rm Ca_{2-x}Sr_xRuO_4$ has been successively driven from an AFM insulator (x $< 0.2$) to AFM correlated metal ($0.2 <$ x $< 0.5$), a nearly FM metal (x $\sim 0.5$), and finally to a non-magnetic Fermi liquid (x $\sim 2$). The dominant effects are only structural modifications due to the larger Sr ionic size, since the substitution is isovalent.\cite{nakatsuji_PRL_2000} With increasing x, the distortion occurs in steps, resulting in removal of first the flattening of the octahedra, then the tilting, and finally the rotation around the $c$ axis \cite{friedt_PRB_2001,fang_PRB_2001,fang_PRB_2004}. Although the substitution is isovalent, only the magnetism of $\rm Ca_{2-x}Sr_{x}RuO_{4}$ is affected in the sequence given above. The bilayer series $\rm (Sr_{1-x}Ca_x)_3 Ru_2 O_7$ also exhibits complex magnetic ground states, ranging from itinerant metamagnet to quasi-2D heavy-mass nearly FM metal, and finally to long-range AFM order. The increase in Ca concentration is associated with a similar tuning of distortion manifested as octahedral rotation, tilting, and flattening.\cite{peng_PRB_2010}

The $\rm Ca_2RuO_4$ compound undergoes a peculiar non-magnetic metal-insulator transition (MIT) at 356 K and a magnetic transition at $T_{\rm N}\approx$ 113 K with observed magnetic moment of $1.3~ \mu_{\rm B}$.\cite{nakatsuji_JPSP_1997,braden_PRB_1998,alexander_PRB_1999,friedt_PRB_2001} The MIT is associated with a structural transition from L-phase (long octahedral $c$-axis) to S-phase (short $c$-axis) due to the continuous flattening of octahedra till the onset of magnetic order at $T_{\rm N}$.\cite{gorelov_PRL_2010} Compared to the isoelectronic member $\rm Sr_2RuO_4$,\cite{nakatsuji_PRL_2000,friedt_PRB_2001} this system has severe structural distortions due to the small $\rm Ca^{2+}$ ionic size, resulting in a compression, rotation, and tilting of the $\rm RuO_6$ octahedra. Thus, the low-temperature phase is characterized by highly distorted
$\rm RuO_6$ octahedra and canted AFM order with moments lying along the crystal $b$ axis.\cite{fang_PRB_2001,Kunkemoller_PRL_2015} Such transitions in $\rm Ca_2RuO_4$ have been identified in temperature,\cite{nakatsuji_PRB_2000} hydrostatic pressure,\cite{nakamura_PRB_2002,steffens_PRB_2005} epitaxial strain,\cite{dietl_APL_2018} chemical substitution,\cite{nakatsuji_PRL_2000,nakamura_PRB_2002,steffens_PRB_2011} and electrical current studies.\cite{nakamura_SREP_2013,okazaki2_JPSJ_2013} 

Earlier investigations of the magnetism in $\rm Ca_2RuO_4$ were based on the $S=1$ local
moment picture corresponding to the approximate electronic configuration $yz^1xz^1xy^2$ in the weak SOC limit. The local moments in the $yz,xz$ orbitals originate from the octahedral compression-induced large tetragonal crystal field ($\approx 0.3$ eV), which lifts the degeneracy of the $t_{\rm 2g}$ states by lowering the $xy$ orbital energy, resulting in nominally doubly occupied $xy$ orbital.\cite{fang_PRB_2004,liebsch_PRL_2007,gorelov_PRL_2010,zhang_PRB_2017} Later, an alternative scenario was proposed in which the magnetism is of the Van Vleck type,\cite{khaliullin_PRL_2013,akbari_PRB_2014} involving the $t_{\rm 2g}^4$ spin-orbit ground state with total angular momentum $J=0$. The proposal of excitonic behavior within this scenario has been applied for the description of magnetic excitations in $\rm Ca_2RuO_4$,
\cite{jain_NATPHY_2017,fatuzzo_PRB_2015,das_PRX_2018,porter_PRB_2018,gretarsson_PRB_2019}
including the putative Higgs-like mode at $\sim 50$ meV reported in recent inelastic neutron scattering studies.\cite{jain_NATPHY_2017,souliou_PRL_2017} However, the $J=0$ scenario is hard to reconcile with recent X-ray scattering, angle-resolved photoemission spectroscopy (ARPES) measurements, and first-principle studies.\cite{fatuzzo_PRB_2015,sutter_NATCOM_2017,friedt_PRB_2001,fang_PRB_2004,liebsch_PRL_2007,gorelov_PRL_2010,zhang_PRB_2017,zhang_PRB_2020}
In this context, it is important to note that in recent theoretical studies based on local-density approximation, dynamical mean-field theory, and many-body perturbation theory, the intrinsic Higgs-like amplitude mode has been shown to be compatible with the first scenario.\cite{zhang_PRB_2020}

The rich phenomenology exhibited by $\rm Ca_2RuO_4$ as discussed above highlights the complex interplay due to intimately intertwined roles of structural distortion, octahedral tilting and rotation, SOC, and Coulomb interaction terms. The fully self consistent approach introduced recently to investigate the magnetic reorientation transition provides the required unified framework.\cite{ruthenate_one_2020} All orbital mixing effects are implicitly included in the orbital off-diagonal spin and charge condensates and therefore treated on the same footing in this approach. For isospin dynamics in the bilayer iridate heterostructure as well, the self consistent approach provides a much more compact formalism compared to phenomenological spin models.\cite{mohapatra_PRB_2019} 

In this paper, we will extend the above approach to investigate subtle magnetic order tuning effects of other structural features such as the octahedral tilting and rotation. We will show that for realistic SOC strength and reduced tetragonal distortion, there exists a tunable regime wherein magnetic order can be tuned (AFM or FM) by octahedral tilting. For small tetragonal distortion, we also find FM order along the crystal $b$ axis as a fully self consistent solution. These results are in agreement with the observed behavior in $\rm Ca_{2-x}Sr_xRuO_4$ and the bilayer ruthenate compound $\rm Ca_3 Ru_2 O_7$, respectively. We will also compare the results obtained using the restricted and fully self consistent calculations in order to highlight the important role of orbital off-diagonal condensates in generating strongly anisotropic SOC renormalization and strongly enhanced orbital magnetic moments. 


The structure of this paper is as follows. After introducing the three-orbital model and Coulomb interaction terms in Sec. II, and reviewing the interaction contributions including orbital off-diagonal spin and charge condensates in Sec. III, the orbital magnetic moment and SOC renormalization are discussed in Sec. IV. Results of the fully self consistent determination of magnetic order are presented in Sec. V. The tunable magnetic regime for reduced tetragonal distortion, stabilization of AFM order with decreasing octahedral tilting, and FM ($b$) order are discussed in Secs. VI and VII. Finally, some conclusions are presented in Sec. VIII. 


\section{Three orbital model and Coulomb interactions}
In the three-orbital ($\mu=yz,xz,xy$), two-spin ($\sigma=\uparrow,\downarrow$) basis defined with respect to a common spin-orbital coordinate system, we consider the Hamiltonian ${\cal H} = {\cal H}_{\rm SOC} + {\cal H}_{\rm cf} + {\cal H}_{\rm band} + {\cal H}_{\rm int}$ within the $t_{\rm 2g}$ manifold. The spin-orbit coupling term ${\cal H}_{\rm SOC}$, which explicitly breaks SU(2) spin rotation symmetry and therefore generates anisotropic magnetic interactions from its interplay with other Hamiltonian terms, is discussed in Appendix A.

For the band and crystal field terms together, we consider:
\begin{eqnarray}
{\cal H}_{\rm band+cf} &=& 
\sum_{{\bf k} \sigma s} \psi_{{\bf k} \sigma s}^{\dagger} \left [ \begin{pmatrix}
{\epsilon_{\bf k} ^{yz}}^\prime & 0 & 0 \\
0 & {\epsilon_{\bf k} ^{xz}}^\prime & 0 \\
0 & 0 & {\epsilon_{\bf k} ^{xy}}^\prime + \epsilon_{xy}\end{pmatrix} \delta_{s s^\prime}
+ \begin{pmatrix}
\epsilon_{\bf k}^{yz} & \epsilon_{\bf k}^{yz|xz} & \epsilon_{\bf k}^{yz|xy} \\
-\epsilon_{\bf k} ^{yz|xz} & \epsilon_{\bf k} ^{xz} & \epsilon_{\bf k}^{xz|xy} \\
-\epsilon_{\bf k}^{yz|xy} & -\epsilon_{\bf k}^{xz|xy} & \epsilon_{\bf k} ^{xy} \end{pmatrix} \delta_{\bar{s} s^\prime } \right] \psi_{{\bf k} \sigma s^\prime} \nonumber \\
\label{three_orb_two_sub}
\end{eqnarray} 
in the composite three-orbital, two-sublattice ($s,s'={\rm A,B}$) basis. Here the energy offset $\epsilon_{xy}$ (relative to the degenerate $yz/xz$ orbitals) represents the tetragonal distortion induced crystal field effect, and the band dispersion terms in the two groups, corresponding to hopping terms connecting the same and opposite sublattice(s), are given by: 
\begin{eqnarray}
\epsilon_{\bf k} ^{xy} &=& -2t_1(\cos{k_x} + \cos{k_y}) \nonumber \\
{\epsilon_{\bf k} ^{xy}} ^{\prime} &=& - 4t_2\cos{k_x}\cos{k_y} - \> 2t_3(\cos{2{k_x}} + \cos{2{k_y}}) \nonumber \\
\epsilon_{\bf k} ^{yz} &=& -2t_5\cos{k_x} -2t_4 \cos{k_y} \nonumber \\
\epsilon_{\bf k} ^{xz} &=& -2t_4\cos{k_x} -2t_5 \cos{k_y}  \nonumber \\
\epsilon_{\bf k} ^{yz|xz} &=&  -2t_{m1}(\cos{k_x} + \cos{k_y}) \nonumber \\
\epsilon_{\bf k} ^{xz|xy} &=&  -2t_{m2}(2\cos{k_x} + \cos{k_y}) \nonumber \\
\epsilon_{\bf k} ^{yz|xy} &=&  -2t_{m3}(\cos{k_x} + 2\cos{k_y}) . 
\label{band}
\end{eqnarray}

\begin{figure}
\vspace*{0mm}
\hspace*{0mm}
\psfig{figure=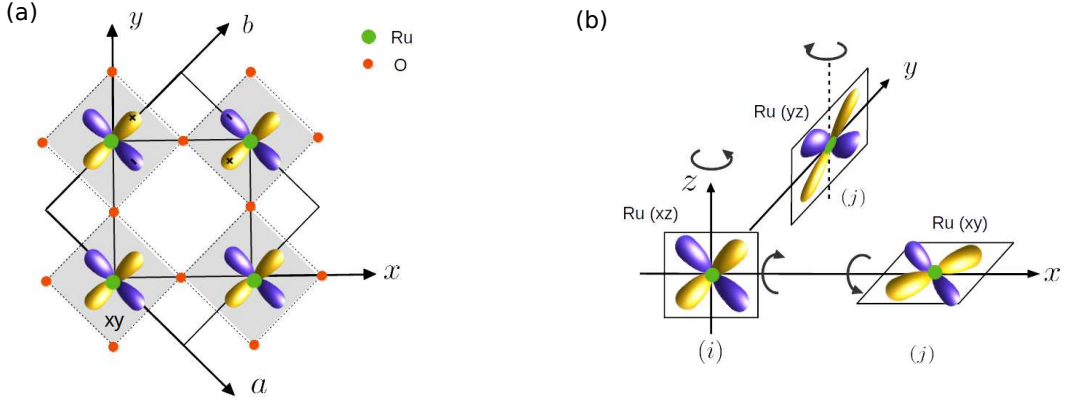,angle=0,width=140mm,angle=0}
\caption{(a) The common spin-orbital coordinate axes ($x-y$) along the Ru-O-Ru directions, shown along with the crystal axes $a,b$. (b) Octahedral rotation about the $z$ axis yields orbital mixing hopping terms between $xz$ and $yz$ orbitals. Octahedral tilting about the crystal $a$ axis is resolved along the $x,y$ axes, resulting in orbital mixing hopping terms between the $xy$ and $xz,yz$ orbitals.} 
\label{axes}
\end{figure}

Here $t_1$, $t_2$, $t_3$ are respectively the first, second, and third neighbor hopping terms for the $xy$ orbital. For the $yz$ ($xz$) orbital, $t_4$ and $t_5$ are the NN hopping terms in $y$ $(x)$ and $x$ $(y)$ directions, respectively, corresponding to $\pi$ and $\delta$ orbital overlaps. Octahedral rotation and tilting induced orbital mixings are represented by the NN hopping terms $t_{m1}$ (between $yz$ and $xz$) and $t_{m2},t_{m3}$ (between $xy$ and $xz,yz$). We have taken hopping parameter values: ($t_1$, $t_2$, $t_3$, $t_4$, $t_5$)=$(-1.0, 0.5, 0, -1.0, 0.2)$, and for the orbital mixing terms: $t_{m1}$=0.2 and $t_{m2}$=$t_{m3}$=0.15 ($\approx 0.2/\sqrt{2}$), all in units of the realistic hopping energy scale $|t_1|$=200meV.\cite{khaliullin_PRL_2013,akbari_PRB_2014,feldmaier_arxiv_2019} 
The choice $t_{m2}=t_{m3}$ corresponds to the octahedral tilting axis oriented along the $-\hat{x}+\hat{y}$ direction, which is equivalent to the crystal $-a$ direction, as shown in Fig. \ref{axes}. The $t_{m1}$ and $t_{m2,m3}$ values taken above approximately correspond to octahedral rotation and tilting angles of about $12^\circ$ ($\approx 0.2$ rad) as reported in experimental studies.\cite{steffens_PRB_2005} 

For the on-site Coulomb interaction terms in the $t_{2g}$ basis ($\mu,\nu=yz,xz,xy$), we consider:
\begin{eqnarray}
{\cal H}_{\rm int} &=& U\sum_{i,\mu}{n_{i\mu\uparrow}n_{i\mu\downarrow}} + U^\prime \sum_{i,\mu < \nu,\sigma} {n_{i\mu\sigma} n_{i\nu\overline{\sigma}}} + (U^\prime - J_{\mathrm H}) \sum_{i,\mu < \nu,\sigma}{n_{i\mu\sigma}n_{i\nu\sigma}} \nonumber\\ 
&+& J_{\mathrm H} \sum_{i,\mu \ne \nu} {a_{i \mu \uparrow}^{\dagger}a_{i \nu\downarrow}^{\dagger}a_{i \mu \downarrow} a_{i \nu \uparrow}} + J_{\mathrm P} \sum_{i,\mu \ne \nu} {a_{i \mu \uparrow}^{\dagger} a_{i \mu\downarrow}^{\dagger}a_{i \nu \downarrow} a_{i \nu \uparrow}} \nonumber\\ 
&=& U\sum_{i,\mu}{n_{i\mu\uparrow}n_{i\mu\downarrow}} + U^{\prime \prime}\sum_{i,\mu<\nu} n_{i\mu} n_{i\nu} - 2J_{\mathrm H} \sum_{i,\mu<\nu} {\bf S}_{i\mu}.{\bf S}_{i\nu} 
+J_{\mathrm P} \sum_{i,\mu \ne \nu} a_{i \mu \uparrow}^{\dagger} a_{i \mu\downarrow}^{\dagger}a_{i \nu \downarrow} a_{i \nu \uparrow} 
\label{h_int}
\end{eqnarray} 
including the intra-orbital $(U)$ and inter-orbital $(U')$ density interaction terms, the Hund's coupling term $(J_{\rm H})$, and the pair hopping interaction term $(J_{\rm P})$, with $U^{\prime\prime} \equiv U^\prime-J_{\rm H}/2=U-5J_{\rm H}/2$ from the spherical symmetry condition $U^\prime=U-2J_{\mathrm H}$. Here $a_{i\mu\sigma}^{\dagger}$ and $a_{i\mu \sigma}$ are the electron creation and annihilation operators for site $i$, orbital $\mu$, spin $\sigma=\uparrow ,\downarrow$, and the density operator $n_{i\mu\sigma}=a_{i\mu\sigma}^\dagger a_{i\mu\sigma}$, total density operator $n_{i\mu}=n_{i\mu\uparrow}+n_{i\mu\downarrow}=\psi_{i\mu}^\dagger \psi_{i\mu}$, and spin density operator ${\bf S}_{i\mu} = \psi_{i\mu}^\dagger ${\boldmath $\sigma$}$ \psi_{i\mu}$, where $\psi_{i\mu}^\dagger=(a_{i\mu\uparrow}^{\dagger} \; a_{i\mu\downarrow}^{\dagger})$. All interaction terms above are SU(2) invariant and thus possess spin rotation symmetry in real-spin space. In the following, we will take $U=8$ in the energy scale unit (200 meV) and $J_{\rm H}=U/5$, so that $U=1.6$eV, $U^{\prime\prime}=U/2=0.8$eV, and $J_{\rm H}=0.32$eV. These are comparable to reported values extracted from RIXS ($J_{\rm H}=0.34$eV) and ARPES ($J_{\rm H}=0.4$eV) studies.\cite{gretarsson_PRB_2019,sutter_NATCOM_2017} 

For moderate tetragonal distortion ($\epsilon_{xy}\approx-1$), the $xy$ orbital in the $4d^4$ compound $\rm Ca_2 Ru O_4$ is nominally doubly occupied and magnetically inactive, while the nominally half-filled and magnetically active $yz,xz$ orbitals yield an effectively two-orbital magnetic system. Hund's coupling between the two $S=1/2$ spins results in low-lying (in-phase) and appreciably gapped (out-of-phase) spin fluctuation modes. The in-phase modes of the $yz,xz$ orbital $S=1/2$ spins correspond to an effective $S=1$ spin system. However, the rich interplay between SOC, Coulomb interaction terms, octahedral tilting and rotation, and tetragonal distortion results in complex magnetic behaviour which crucially involves the $xy$ orbital, and is therefore beyond the above simplistic picture. 

\begin{figure}
\vspace*{0mm}
\hspace*{0mm}
\psfig{figure=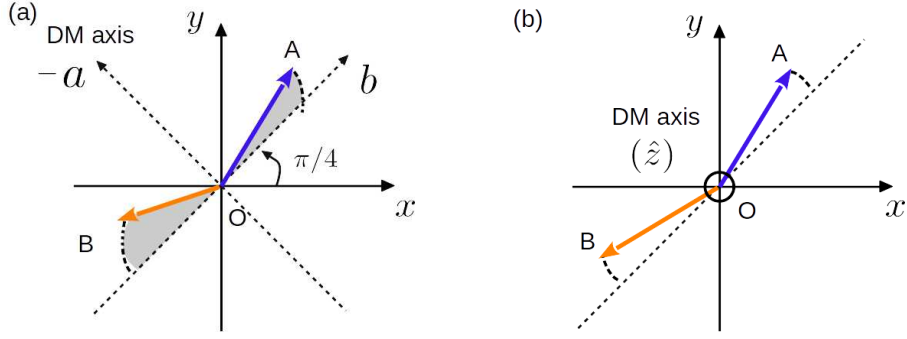,angle=0,width=120mm,angle=0}
\caption{AFM ordering with SOC induced easy plane anisotropy, along with spin cantings about the (a) crystal $a$ axis and (b) crystal $c$ axis, due to the effective Dzyaloshinski-Moriya (DM) interactions induced by the staggered octahedral tilting and rotation, respectively. Octahedral tilting about the crystal $a$ axis yields the magnetic easy axis along the crystal $b$ (perpendicular) direction.} 
\label{canting}
\end{figure}

Some of the important physical consequences of the interplay between different physical elements, such as SOC induced easy plane anisotropy, octahedral tilting induced easy axis anisotropy, and tilting and rotation induced spin cantings, are discussed in terms of induced anisotropic spin interactions in Appendices A and B. The resulting magnetic order for moderate tetragonal distortion is depicted in Fig. \ref{canting}. However, for reduced tetragonal distortion, the coupled spin-orbital fluctuations become important, and the appropriate framework is provided by the self-consistent approach which is briefly reviewed below. 


\section{Coulomb interaction contributions including orbital off-diagonal spin and charge condensates}
We consider the various Coulomb interaction terms (Eq. \ref{h_int}) in the Hartree-Fock (HF) approximation, including contributions from the orbital off-diagonal (OOD) spin and charge condensates (Appendix C), which play an important role in the self consistent determination of magnetic order. The resulting local spin and charge terms can be written as:
\begin{equation}
[{\cal H}_{\rm int}^{\rm HF}] = [{\cal H}_{\rm int}^{\rm HF}]_{\rm normal} + [{\cal H}_{\rm int}^{\rm HF}]_{\rm OOD} = \sum_{i\mu\nu} \psi_{i\mu}^{\dagger} \left [
-\makebox{\boldmath $\sigma . \Delta$}_{i\mu\nu} + {\cal E}_{i\mu\nu} {\bf 1} \right ] \psi_{i\nu} 
\label{h_hf} 
\end{equation}  
where the normal ($\mu=\nu$) spin and charge fields are self-consistently determined from:
\begin{eqnarray}
2\Delta_{i\mu}^\alpha &=& U\langle \sigma_{i\mu}^\alpha \rangle + J_{\rm H} \sum_{\nu < \mu} \langle \sigma_{i\nu}^\alpha \rangle \;\;\;\;\;(\alpha=x,y,z) \nonumber \\
{\cal E}_{i\mu} &=& \frac{U\langle n_{i\mu}\rangle}{2} + U'' \sum_{\nu < \mu} \langle n_{i\nu} \rangle 
\label{selfcon}
\end{eqnarray}
in terms of the local charge density $\langle n_{i\mu}\rangle$ and spin density components $\langle \sigma_{i\mu}^\alpha \rangle$. For $\langle n_{yz}\rangle=\langle n_{xz}\rangle$, the Coulomb renormalized tetragonal splitting is obtained as:
\begin{eqnarray}
\tilde{\delta}_{\rm tet} &=& \tilde{\epsilon}_{xz,yz} - \tilde{\epsilon}_{xy} = (\epsilon_{xz,yz} - \epsilon_{xy}) 
+ \left [{\cal E}_{yz,xz} - {\cal E}_{xy} \right ] \nonumber \\
& = & \delta_{\rm tet} + \left [\frac{U\langle n_{yz,xz}\rangle}{2} + U'' \langle n_{yz,xz} + n_{xy}\rangle \right ] - \left [\frac{U\langle n_{xy}\rangle}{2} + 2U'' \langle n_{yz,xz}\rangle \right ] \nonumber \\
& = & \delta_{\rm tet} + (U'' -U/2)\langle n_{xy} - n_{yz,xz}\rangle 
\end{eqnarray}
which shows that the Coulomb renormalization identically vanishes for the realistic relationship $U''=U/2$ for $4d$ orbitals, as discussed in Sec. II. 

Similarly, the orbital off-diagonal fields are self-consistently determined from:
\begin{eqnarray}
\makebox{\boldmath $\Delta$}_{i\mu\nu} &=& \left (\frac{U''}{2} + \frac{J_{\rm H}}{4} \right ) \langle \makebox{\boldmath $\sigma$}_{i\nu\mu} \rangle + \left (\frac{J_{\rm P}}{2} \right ) \langle \makebox{\boldmath $\sigma$}_{i\mu\nu} \rangle \nonumber \\
{\cal E}_{i\mu\nu} &=& \left (-\frac{U''}{2} + \frac{3J_{\rm H}}{4} \right ) \langle n_{i\nu\mu} \rangle + \left (\frac{J_{\rm P}}{2}\right ) \langle n_{i\mu\nu} \rangle  
\label{sc_od}
\end{eqnarray}
in terms of the corresponding condensates $\langle \makebox{\boldmath $\sigma$}_{i\nu\mu} \rangle \equiv \langle \psi_{i\nu}^{\dagger} \makebox{\boldmath $\sigma$} \psi_{i\mu} \rangle$ and $\langle n_{i\nu\mu} \rangle \equiv \langle \psi_{i\nu}^{\dagger} {\bf 1} \psi_{i\mu} \rangle$. For each orbital pair ($\mu,\nu$) = ($yz,xz$), ($xz,xy$), ($xy,yz$), there are three components ($\alpha=x,y,z$) for the spin condensates $\langle \psi_\mu^\dagger \sigma_\alpha \psi_\nu \rangle$ and one charge condensate $\langle \psi_\mu^\dagger {\bf 1} \psi_\nu \rangle$. This is analogous to the three-plus-one normal spin and charge condensates for each of the three orbitals $\mu=yz,xz,xy$. Before proceeding with the fully self consistent calculation, we first discuss how the most dominant off-diagonal condensates ${\rm Im} \langle \psi_\mu^\dagger \psi_\nu \rangle$ and ${\rm Im} \langle \psi_\mu ^\dagger \sigma_\alpha \psi_\nu \rangle$ result in coupling of orbital moments to orbital fields and interaction induced SOC renormalization. 

\section{Orbital magnetic moment and SOC renormalization}
The off-diagonal charge condensates $\langle \psi_\mu^\dagger \psi_\nu \rangle$ directly yield the orbital magnetic moments:
\begin{eqnarray}
\langle L_x \rangle &=& \langle \psi_{xz}^\dagger (-i) \psi_{xy} \rangle + \langle \psi_{xy}^\dagger (i) \psi_{xz} \rangle \nonumber \\
&=& -i \langle \psi_{xz}^\dagger \psi_{xy} \rangle + i \langle \psi_{xz}^\dagger \psi_{xy} \rangle^* = 2 {\rm Im} \langle \psi_{xz}^\dagger \psi_{xy} \rangle
\end{eqnarray}
and similarly for the other components. Accordingly, the charge term in Eq. (\ref{h_hf}), of which only the anti-symmetric part is non-vanishing (see Appendix C), can be represented as a coupling of orbital angular momentum operators to orbital fields:
\begin{eqnarray}
[{\cal H}_{\rm int}^{\rm HF}]_{\rm OOD}^{\rm charge}(i)|_{\rm anti-sym} &=& 
-\frac{U''_{\rm c|a}}{2} \sum_{\mu < \nu} \langle n_{\mu\nu}\rangle^{\rm Im} \left [ \psi_\mu^\dagger (-i) \psi_\nu + {\rm H.c.} \right ] \nonumber \\
&=& -\frac{U''_{\rm c|a}}{4} \left [\langle L_x\rangle L_x + \langle L_y\rangle L_y + \langle L_z\rangle L_z \right ]
\label{orb_int}
\end{eqnarray}
which corresponds to an effective isotropic interaction $-(U''_{\rm c|a}/8){\bf L}.{\bf L}$ between orbital moments, and will therefore enhance the $\langle L_\alpha\rangle$ values in the HF calculation.  

Similarly, for the spin term in Eq. (\ref{h_hf}), the anti-symmetric part (see Appendix C) can be represented in terms of the spin-orbital operators:
\begin{eqnarray}
[{\cal H}_{\rm int}^{\rm HF}]_{\rm OOD}^{\rm spin} (i)|_{\rm anti-sym} & = & -(U''_{\rm s|a}/2) \sum_{\mu < \nu} \langle \makebox{\boldmath $\sigma$}_{\mu\nu} \rangle^{\rm Im} . \left [ \psi_\mu ^\dagger (-i \makebox{\boldmath $\sigma$}) \psi_\nu + {\rm H.c.} \right ] \nonumber \\
& = & - \sum_{\alpha=x,y,z} \left [ \lambda_\alpha ^{\rm int} L_\alpha S_\alpha + \sum_{\beta\ne\alpha} \lambda_{\alpha\beta}^{\rm int} L_\alpha S_\beta \right ]
\label{soc_both}
\end{eqnarray}
where the interaction-induced SOC renormalization terms:
\begin{equation}
\lambda_\alpha ^{\rm int} = U''_{\rm s|a} {\rm Im} \langle \psi_\mu^\dagger \sigma_\alpha \psi_\nu \rangle = U''_{\rm s|a} \langle \psi_\mu^\dagger (-i \sigma_\alpha ) \psi_\nu \rangle^{\rm Re} = U''_{\rm s|a} \langle L_\alpha S_\alpha \rangle   
\label{soc_ren} 
\end{equation}
for the orbital pair $\mu,\nu$ corresponding to component $\alpha$. Although the off-diagonal SOC terms $(L_\alpha S_\beta)$ are smaller than the diagonal terms ($\lambda_{\alpha\beta}^{\rm int} < \lambda_\alpha^{\rm int}$), they are still significant as shown in the next section. 

\begin{table}[t] 
\caption{Off-diagonal spin and charge condensates evaluated for the three orbital pairs in the restricted self consistent state. Here the effective SOC = 2.0, $\epsilon_{xy}=0$, and $t_{m1,m2,m3}=0$.} 
\centering 
\begin{tabular}{l r r r r} \\  
\hline\hline 
Orbital pair & $\langle \psi_\mu^\dagger \sigma_x \psi_\nu \rangle$ & $\langle \psi_\mu^\dagger \sigma_y \psi_\nu \rangle$ & $\langle \psi_\mu^\dagger \sigma_z \psi_\nu \rangle$ & $\langle \psi_\mu^\dagger {\bf 1} \psi_\nu \rangle$ \\ [0.5ex]
\hline 
$yz-xz$ (A) & (0.046,0) & (0.046,0) & (0,0.236) & ($-$0.066,0) \\ [1ex]
$xz-xy$ (A) & (0,0.313) & (0,0.066) & (0.075,0) & (0,$-$0.220) \\ [1ex]
$xy-yz$ (A) & (0,0.066) & (0,0.313) & (0.075,0) & (0,$-$0.220) \\[1ex] \hline
\end{tabular}
\label{table1new}
\end{table}

\section{Self-consistent determination of magnetic order}
\begin{figure}
\vspace*{-0mm}
\hspace*{0mm}
\psfig{figure=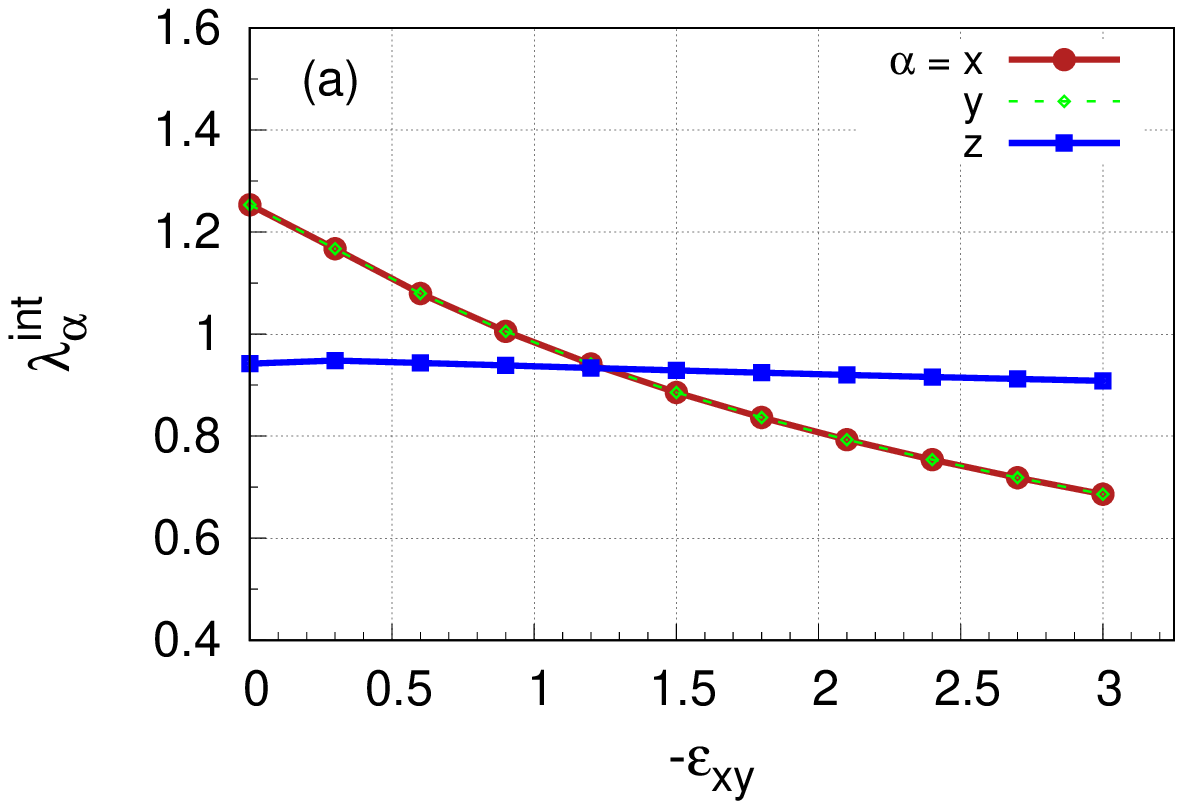,angle=0,width=80mm}
\psfig{figure=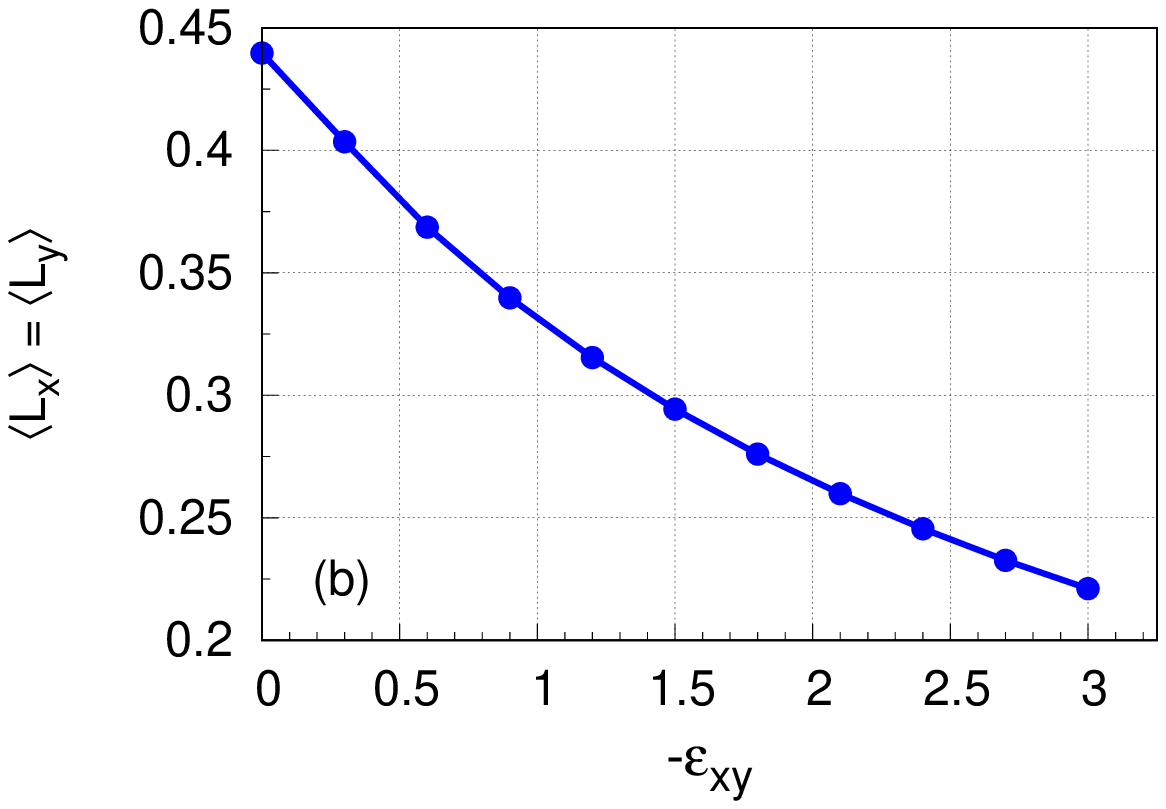,angle=0,width=80mm}
\caption{Variation of the (a) SOC renormalization $\lambda^{\rm int}_\alpha \approx U'' {\rm Im} \langle \psi_\mu ^\dagger \sigma_\alpha \psi_\nu \rangle$ and (b) orbital magnetic moment $\langle L_\alpha \rangle = 2{\rm Im} \langle \psi_\mu^\dagger \psi_\nu \rangle$, calculated from the orbital off-diagonal spin and charge condensates, with tetragonal splitting $\delta_{\rm tet} \equiv \epsilon_{yz,xz} - \epsilon_{xy}$. Here $U=8$ and effective SOC $\lambda^{\rm eff}=2$.} 
\label{anom}
\end{figure}

It is instructive to first evaluate the magnitude of the orbital off-diagonal condensates in the restricted self consistent state of $[{\cal H}_{\rm SOC}] + [{\cal H}_{\rm band}] + [{\cal H}_{\rm int}^{\rm HF}]_{\rm normal}$. Calculated results for these condensates are given in Table \ref{table1new} for AFM order along the crystal $b$ direction. The dominant off-diagonal condensates are seen to be ${\rm Im} \langle \psi_\mu^\dagger \psi_\nu \rangle$ and ${\rm Im} \langle \psi_\mu ^\dagger \sigma_\alpha \psi_\nu \rangle$, where the orbital pairs $\mu,\nu$ correspond to the components $\alpha=x,y,z$ as in Eq. (\ref{soc}). Fig. \ref{anom} shows the behavior of related physical quantities, reflecting the reduced mixing between $xy$ and $yz,xz$ orbitals with increasing tetragonal distortion. The octahedral tilting and rotation have been neglected here for simplicity. As seen in Fig. \ref{anom}(a), for moderate tetragonal distortion ($-\epsilon_{xy} \approx 1.0$), the SOC renormalization is close to 1 for all three components $\alpha=x,y,z$. As the effective SOC strength $\lambda^{\rm eff}=\lambda+\lambda^{\rm int}$ was taken as 2, this calculation is approximately self consistent for the bare SOC strength $\lambda \approx 1$.  

\begin{table}[b] \vspace*{0mm}
\caption{Self consistently determined magnetization and density values for the three orbitals ($\mu$) on the two sublattices ($s$), including the octahedral rotation and tilting.} 
\centering 
\begin{tabular}{l c c c c} \\  
\hline\hline 
$\mu$ (s) & $m_\mu^x$ & $m_\mu^y$ & $m_\mu^z$ & $n_\mu$ \\ [0.5ex]
\hline 
$yz$ (A) & 0.533 & 0.594 & 0.101 & 1.133 \\[1ex]
$xz$ (A) & 0.518 & 0.639 & 0.111 & 1.111 \\[1ex]
$xy$ (A) & 0.095 & 0.128 & 0.055 & 1.756 \\[1ex] \hline
\end{tabular} \hspace{10mm}
\begin{tabular}{l c c c c} \\  
\hline\hline 
$\mu$ (s) & $m_\mu^x$ & $m_\mu^y$ & $m_\mu^z$ & $n_\mu$ \\ [0.5ex]
\hline 
$yz$ (B) & $-$0.639 & $-$0.518 & 0.111 & 1.111 \\[1ex]
$xz$ (B) & $-$0.594 & $-$0.533 & 0.101 & 1.133 \\[1ex]
$xy$ (B) & $-$0.128 & $-$0.095 & 0.055 & 1.756 \\[1ex] \hline 
\end{tabular}
\label{table1}
\end{table}

A fully self consistent treatment is evidently required in view of the strong magnitude of the orbital off-diagonal spin and charge condensates, and we now discuss results of this calculation. The magnetization and density values for the three orbitals are presented in Table \ref{table1}, all off-diagonal spin and charge condensates in Table \ref{table2}, and the renormalized SOC values and orbital magnetic moments in Table \ref{table3}. Here $U=8$, $\epsilon_{xy}=-1.0$, the bare SOC value $\lambda=1$, and the staggered octahedral rotation ($t_{m1}=0.2$) and tilting ($t_{m2}=t_{m3}=0.15$) have been included. The dominant $yz,xz$ moments (Table \ref{table1}) show the expected AFM order along the crystal $b$ easy axis, along with spin cantings in and about the $z$ direction due to the octahedral tilting and rotation (Appendices A and B). The strongly anisotropic SOC renormalization and strongly enhanced orbital moments (Table IV), as well as the significant magnitude of the off-diagonal SOC terms highlight the important role of orbital off-diagonal condensates. For the off-diagonal SOC term $(L_\alpha S_\beta)$ in Eq. \ref{soc_both}, from Table \ref{table2} we obtain $\lambda_{xy}^{\rm int} \approx U'' \times 0.111 \approx 0.4$ on the A sublattice, whereas the bare SOC = 1.0. 

\begin{table} \vspace*{0mm}
\caption{Self consistently determined off-diagonal spin and charge condensates for the three orbital pairs on the two sublattices.} 
\centering 
\begin{tabular}{l r r r r} \\  
\hline\hline 
Orbital pair & $\langle \psi_\mu^\dagger \sigma_x \psi_\nu \rangle$ & $\langle \psi_\mu^\dagger \sigma_y \psi_\nu \rangle$ & $\langle \psi_\mu^\dagger \sigma_z \psi_\nu \rangle$ & $\langle \psi_\mu^\dagger {\bf 1} \psi_\nu \rangle$ \\ [0.5ex]
\hline 
$yz-xz$ (A) & (0.054,0.019) & (0.057,0.012) & (0.013,0.164) & $-$(0.071,0.041) \\ [1ex]
$xz-xy$ (A) & (0.023,0.279) & (0.037,0.111) & (0.077,0.022) & $-$(0.039,0.238) \\ [1ex]
$xy-yz$ (A) & (0.032,0.101) & (0.043,0.308) & (0.081,0.015) & $-$(0.056,0.260) \\[1ex] \hline
$yz-xz$ (B) & $-$(0.057,0.012) & $-$(0.054,0.019) & (0.013,0.164) & $-$(0.071,0.041) \\[1ex]
$xz-xy$ (B) & (0.043,0.308) & (0.032,0.101) & $-$(0.081,0.015) & (0.056,0.260) \\[1ex]
$xy-yz$ (B) & (0.037,0.111) & (0.023,0.279) & $-$(0.077,0.022) & (0.039,0.238) \\[1ex]\hline 
\end{tabular}
\label{table2}
\end{table}

\begin{table}[b]
\caption{Self consistently determined renormalized SOC values $\lambda_\alpha = \lambda+\lambda_\alpha^{\rm int}$ and the orbital magnetic moments $\langle L_\alpha \rangle$ for $\alpha=x,y,z$ on the two sublattices. Bare SOC $=1.0$} 
\centering 
\begin{tabular}{l r r r r r r} \\  
\hline\hline 
$s$ & $\lambda_x$ & $\lambda_y$ & $\lambda_z$ & $\langle L_x \rangle$ & $\langle L_y \rangle$ & $\langle L_z \rangle$ \\ [0.5ex]
\hline 
A & 1.892 & 1.986 & 1.526 & $-$0.476 & $-$0.520 & $-$0.082 \\[1ex]
B & 1.986 & 1.892 & 1.526 & 0.520 & 0.476 & $-$0.082 \\[1ex]
\hline 
\end{tabular}
\label{table3}
\end{table}

\begin{figure}[t]
\vspace*{-0mm}
\hspace*{0mm}
\psfig{figure=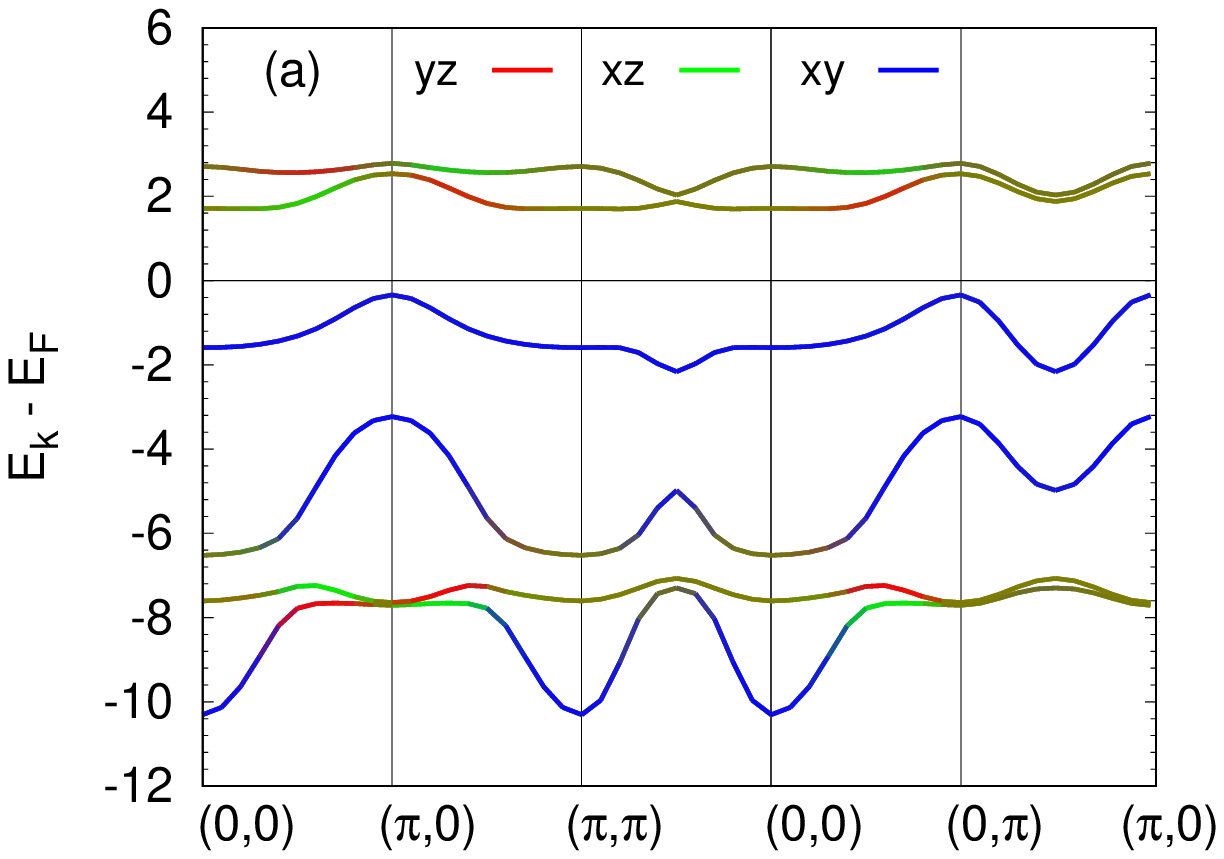,angle=0,width=80mm,angle=0}
\psfig{figure=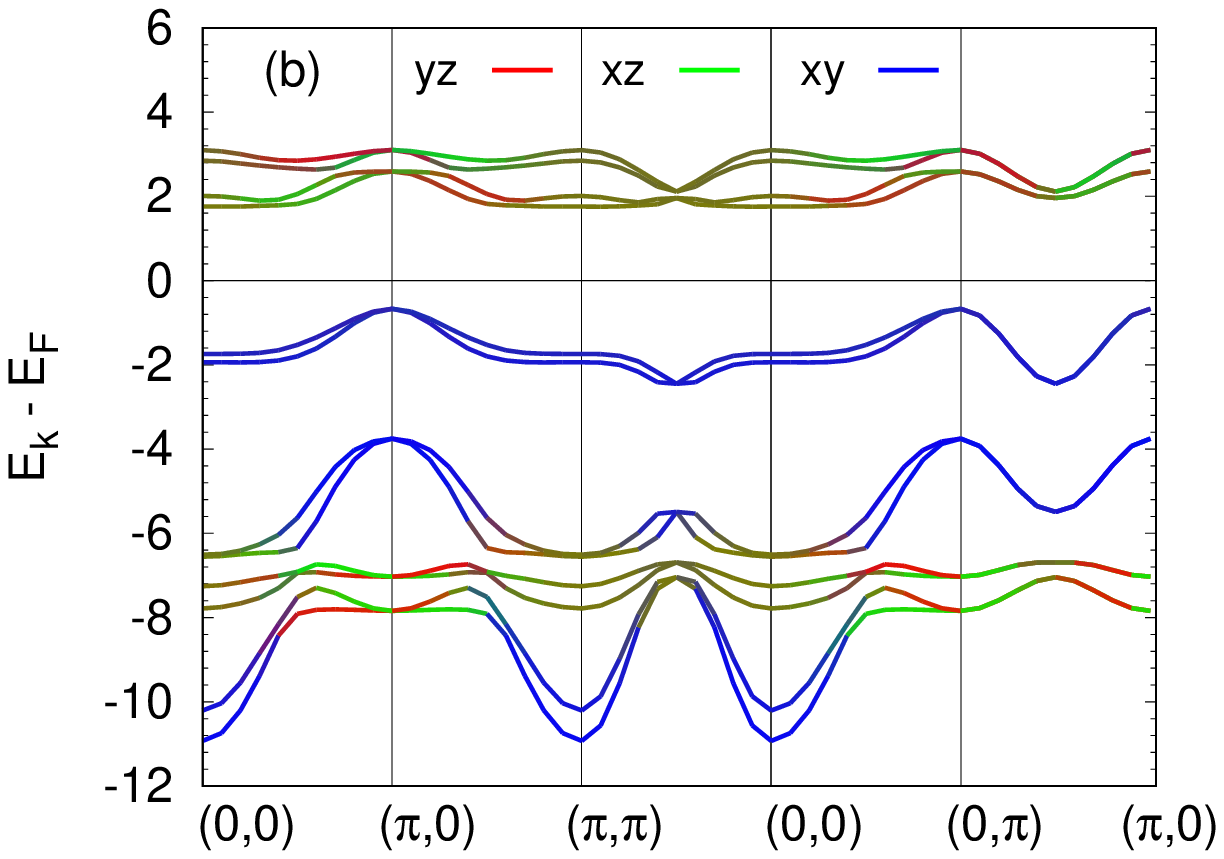,angle=0,width=80mm,angle=0}
\caption{Calculated electronic band structure in the self-consistent AFM state for moderate tetragonal distortion: (a) without and (b) with all off-diagonal spin and charge condensates included, along with octahedral tilting and rotation. Colors indicate dominant orbital weight: red ($yz$), green ($xz$), blue ($xy$). Here $U$=8, $\epsilon_{xy}$=$-$1.0, effective SOC=2.0 in (a) and bare SOC=1.0 in (b).} 
\label{band_all}
\end{figure}

Fig. \ref{band_all} shows the orbital resolved electronic band structure in the self consistent AFM state calculated for the two cases: (a) including only normal condensates and effective SOC, and (b) including all off-diagonal spin and charge condensates along with octahedral rotation and tilting. The band structure shows the narrow AFM sub bands for the magnetically active $yz,xz$ orbitals above and below the Fermi energy due to the dominant exchange field splitting. The relatively smaller splitting between the $xy$ sub bands (both below $E_{\rm F}$) is due to the weaker effect of $yz,xz$ moments through the Hund's coupling. The octahedral tilting and rotation are seen to introduce fine splittings due to the orbital mixing hopping terms. 

Comparison of Figs. \ref{band_all} (a) and (b) shows that the broad features such as energy and dispersion of bands (related to spin and charge densities, SOC, and band terms) are approximately captured in the restricted self consistent calculation (a). However, the strong orbital mixing seen in the band structure (b) reflects the important role of orbital off-diagonal spin and charge condensates in the fully self consistent calculation.

We summarize here the results obtained above for moderate tetragonal distortion ($\epsilon_{xy} \sim -1.0$), with all orbital off-diagonal spin and charge condensates included in the self consistent calculation. With nearly half filled $yz,xz$ orbitals and nearly filled $xy$ orbital, the AFM insulating state is characterized by AFM order due to dominantly $yz,xz$ moments in the SOC induced easy ($a$-$b$) plane and aligned along the octahedral tilting induced easy ($b$) axis, with small canting of moments in and about the crystal $c$ axis. The Coulomb renormalization incorporated by including the orbital off-diagonal condensates leads to strongly enhanced orbital magnetic moments $\langle L_x \rangle$ and $\langle L_y \rangle$ and strongly anisotropic renormalized SOC values ($\lambda_x,\lambda_y > \lambda_z$), as seen by comparing results of the restricted and fully self consistent calculations. The spin cantings become negligible when octahedral tilting and rotation are set to zero. Spin canting in the $c$ direction has been recently observed in resonant elastic X-ray scattering experiments.\cite{porter_PRB_2018}

\section{Stabilization of AFM order by decreasing octahedral tilting --- tunable magnetic order}
For moderate tetragonal distortion ($\epsilon_{xy} \sim -1$), planar AFM order with small $c$ axis canting is obtained in the fully self consistent calculation with octahedral tilting and rotation included. However, with decreasing tetragonal distortion, a sharp magnetic reorientation transition from the dominantly $a-b$ plane AFM order to $c$ axis FM order was obtained recently.\cite{ruthenate_one_2020} With respect to orbital averaged magnetic orders defined as:
\begin{eqnarray}
m_{\rm AFM}^{x-y} &=& (1/3)\sum_\mu \left [ \left (\frac{m_\mu^x (A) - m_\mu^x (B)}{2}\right )^2 +  \left (\frac{m_\mu^y (A) - m_\mu^y (B)}{2}\right )^2 \right ] ^{1/2} \nonumber \\
m_{\rm FM}^z &=& (1/3)\sum_\mu m_\mu^z
\end{eqnarray}
the planar AFM order was found to decrease sharply across the transition, while the FM ($z$) order increases sharply. The electronic state was found to remain insulating down to $\epsilon_{xy}=0$, with filling $n=4$. AFM correlations were seen to persist even after the transition to FM ($z$) order. The transition is obtained only when the off-diagonal condensates are included.

\begin{figure}[t]
\vspace*{0mm}
\hspace*{0mm}
\psfig{figure=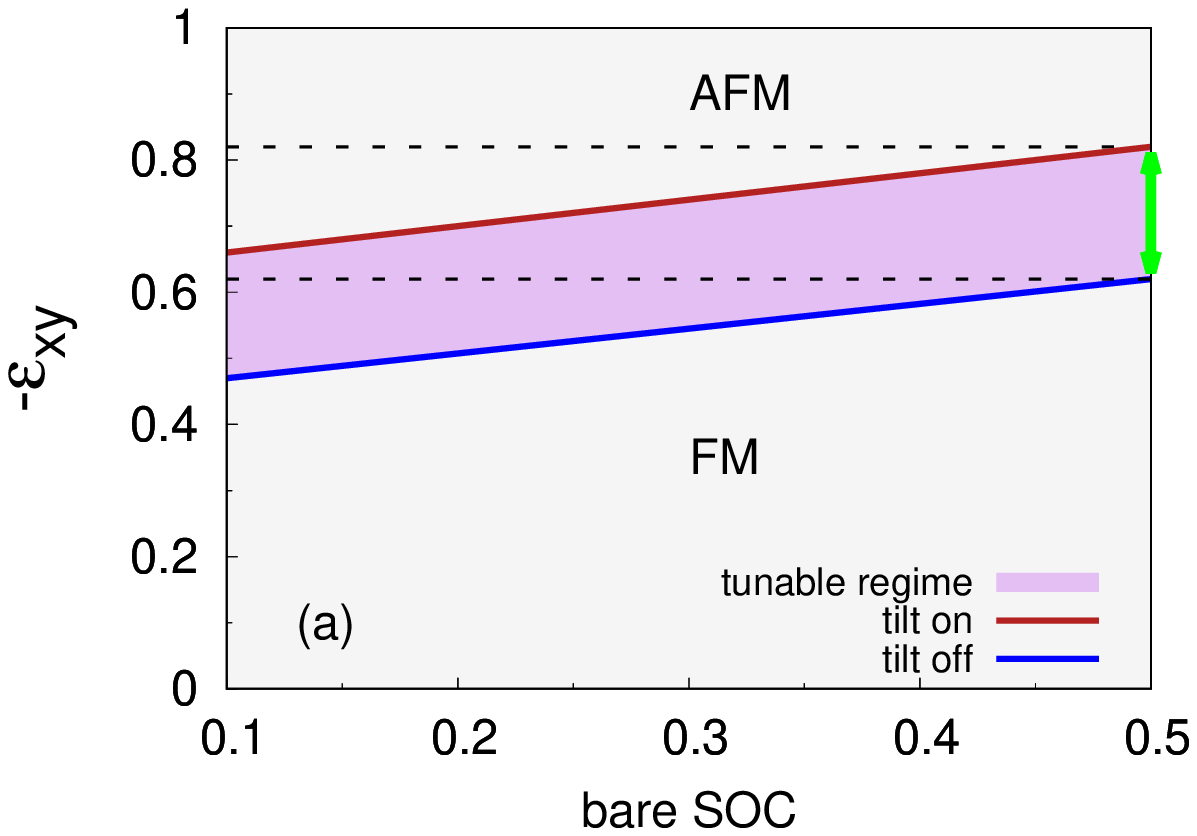,angle=0,width=80mm,angle=0} 
\psfig{figure=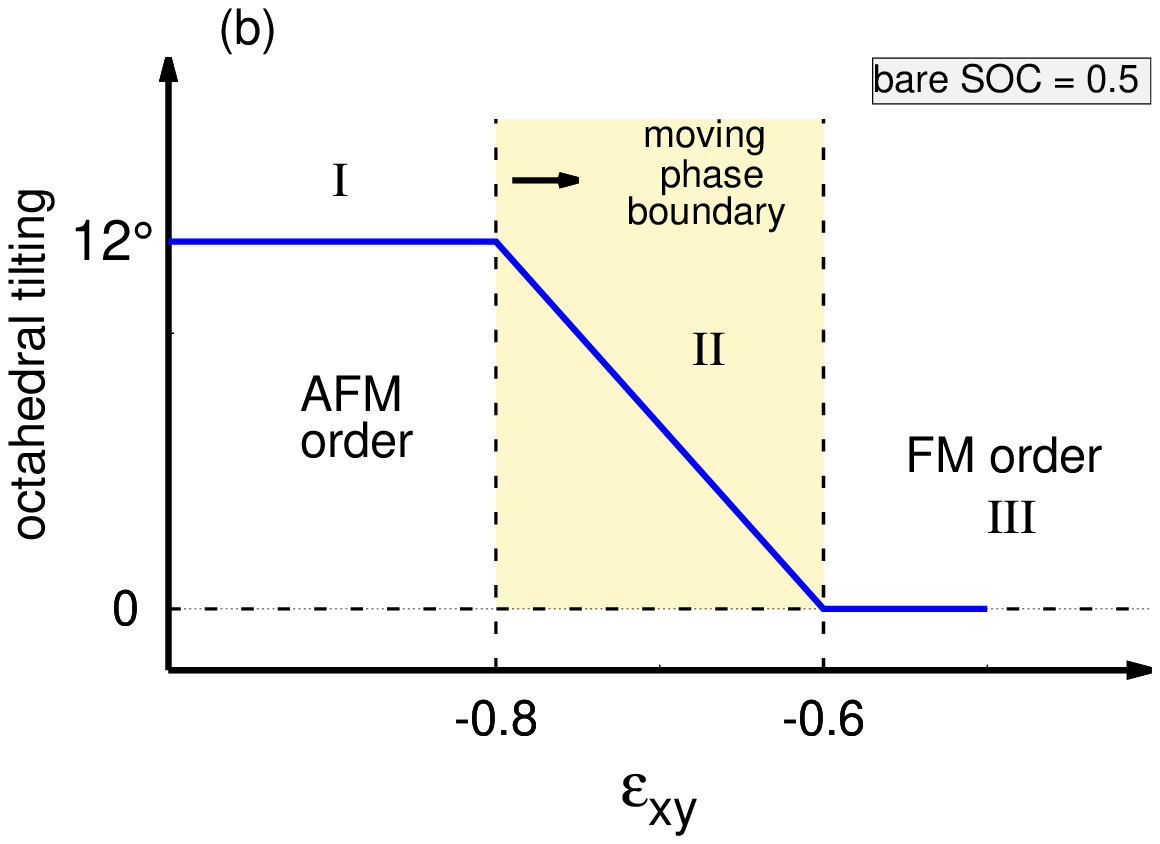,angle=0,width=80mm,angle=0} 
\caption{(a) Magnetic phase boundary between AFM and FM orders in the small SOC regime with and without the octahedral tilting included, showing that AFM order is stabilized by decreasing octahedral tilting. (b) For fixed SOC and with decreasing tetragonal distortion, the AFM order is maintained in the tunable regime (II) by progressively decreasing the octahedral tilting.} 
\label{tilting}
\end{figure}

In the following, we will focus on the effects of octahedral tilting and rotation on the stability of AFM order in the small SOC regime. The magnetic phase boundary between AFM and FM orders is shown in Fig. \ref{tilting} for the two cases: (i) with and (ii) without the octahedral tilting included. The octahedral rotation was found to affect the reorientation transition only weakly, and was therefore retained here for simplicity. Fig. \ref{tilting}(a) shows that the AFM order is stabilized by decreasing octahedral tilting. Within the tunable regime, the magnetic order can be tuned (AFM or FM) by the octahedral tilting magnitude. 

With decreasing octahedral tilting, the NN hopping term $|t_4|$ for the $yz,xz$ orbitals will increase slightly, which will stabilize the AFM order due to enhanced interaction energy ($\sim 4t_4 ^2/U$). Therefore, for fixed SOC and with decreasing tetragonal distortion ($|\epsilon_{xy}|$), the AFM order will be maintained in the tunable regime (II) by progressively decreasing the octahedral tilting, as illustrated in Fig. \ref{tilting}(b). This picture is in agreement with the finding that with increasing x in the isoelectronic series $\rm Ca_{2-x}Sr_x RuO_4$, the structural changes occur in steps, with removal of first the flattening of the octahedra (decreasing tetragonal distortion), then the tilting, and finally the rotation around the $c$ axis.\cite{friedt_PRB_2001,fang_PRB_2001,fang_PRB_2004}

\begin{figure}[t]
\vspace*{-10mm}
\hspace*{0mm}
\psfig{figure=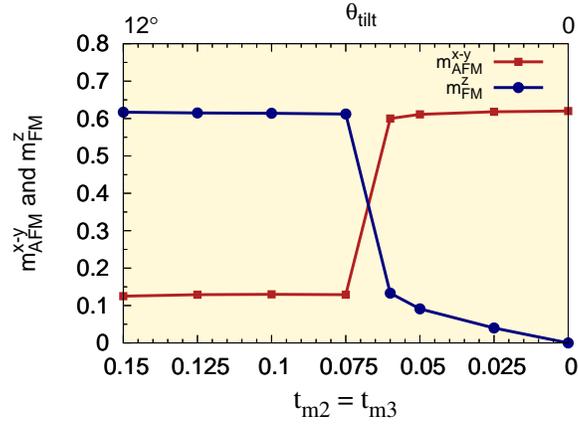,angle=0,width=80mm,angle=0} 
\caption{Stabilization of AFM order with decreasing octahedral tilting.} 
\label{stabilization}
\end{figure}

Fig. \ref{stabilization} shows the stabilization of AFM order with decreasing octahedral tilting, which is equivalent to decreasing orbital mixing hopping terms $t_{m2,m3}$. Here, we have taken $\epsilon_{xy}=-0.7$ corresponding to the midpoint of the tunable regime (Fig. \ref{tilting}) at bare SOC = 0.5, which corresponds to the realistic SOC value of 100 meV. Therefore, as expected from linear behavior, the transition from FM to AFM order occurs near the middle. Although FM order is obtained at the midpoint ($t_{m2}=0.075$), AFM order becomes stable here when $|t_4|$ is increased slightly from 1.0 to 1.1 to incorporate the improved orbital overlap with decreasing tilting as discussed above.  

Fig. \ref{tilting}(a) shows that the tetragonal distortion driven reorientation transition is weakly tuned by SOC and octahedral tilting. The stabilization of AFM order by decreasing octahedral tilting can be understood in terms of an effective $|\epsilon_{xy}|$. The octahedral tilting induces inter-site orbital mixing between $xy$ and $xz,yz$ orbitals ($t_{m2}$ and $t_{m3}$), which effectively reduces $|\epsilon_{xy}|$ due to orbital dependent energy shifts. Decreasing octahedral tilting has the opposite effect and thus enhances the effective $|\epsilon_{xy}|$, which stabilizes the AFM order. 

Finally, we consider the special case of bare SOC = 0. In this case, Im$\langle \psi_\mu^\dagger \psi_\nu \rangle$ and Im$\langle \psi_\mu^\dagger \sigma_\alpha \psi_\nu \rangle$ are identically zero. However, the reorientation transition still occurs with decreasing $|\epsilon_{xy}|$, which is ascribed to finite real parts of the orbital off-diagonal condensates, resulting from octahedral tilting and rotation induced orbital mixing. It should be noted that since there is no SU(2) spin rotation symmetry breaking for SOC = 0, there is no magnetic anisotropy. The above discussion illustrates how all elements get treated on the same footing in the self consistent approach by including the orbital off-diagonal condensates. The reorientation transition occurs irrespective of whether the orbital mixing originates from octahedral tilting or finite SOC.


We have shown above how SOC and octahedral tilting weakly tune the magnetic reorientation transition from AFM to FM order driven by decreasing tetragonal distortion ($\epsilon_{xy}$) due to octahedral de-flattening. The theory presented above provides a microscopic account of magnetic order tuning by structural distortions for fixed SOC value within the tunable regime, as experimentally induced in $\rm Ca_2RuO_4$ by a variety of agents such as hydrostatic pressure, chemical substitution, epitaxial strain due to different substrates, and electrical current. Similar behaviour is realized in the recently synthesized $\rm Ca_2RuO_4$ nanofilm crystal which exhibits robust FM order with enhanced $T_{\rm C}$=180 K due to reduced structural distortions.\cite{nobukane_SREP_2020} Tuning the film thickness to the nanometer range releases the octahedral distortion similar to pressurization effect in the bulk compound, leading to transition from AFM insulating phase to FM metallic phase. With decreasing number of layers, first-principle calculations show decreasing octahedral tilting angle and increasing $c$ axis distance.

\section{FM order in crystal $b$ direction}
Another illustration of magnetic order tuning is provided by the bilayer ruthenate compound ($\rm Ca_3Ru_2O_7$) which exhibits FM order within bilayers due to negligible octahedral flattening.\cite{yoshida_PRB_2005} It undergoes a magnetic transition at $T_{\rm N} = 56$ K and an electronic transition at 48 K with similar $c$-axis compressive transition like the monolayer counterpart, although the effect is much smaller (only $\sim 10\%$ of $\rm Ca_2RuO_4$).\cite{cao_PRB_2003,yoshida_PRB_2004} The octahedra are both rotated and tilted by angles $\phi=15^\circ$, $\theta=13.6^\circ$, respectively.\cite{peng_PRB_2010} The ferromagnetically aligned moments within bilayers lie along the crystal $b$ axis,\cite{mccall_PRB_2003,yoshida_PRB_2004} but the moments are coupled antiferromagnetically between the bilayers, leading to A-type AFM order.\cite{yoshida_PRB_2005} The bilayer ruthenate has been subject of intense investigations due to multitude of interesting low-temperature properties, such as spin-valve and colossal magnetoresistance effects.\cite{lin_PRL_2005,bao_PRL_2008,zhu_PRL_2016} 

As discussed in the previous section, octahedral tilting induced spin canting in $z$ direction [Fig. \ref{canting}(a)] leads to FM ($z$) order for reduced tetragonal distortion. We will now investigate whether FM order in $x-y$ plane can similarly be obtained due to octahedral rotation induced in-plane spin canting [Fig. \ref{canting}(b)]. For this purpose, we searched for fully self consistent solution with purely in-plane magnetic order by starting with in-plane AFM order in different directions. We find that for realistic SOC and reduced tetragonal distortion, the only such self consistent solution corresponds to FM ($b$) order. There is an accompanying weak AFM ($a$) order along the octahedral tilting induced DM axis [Fig. \ref{canting}(a)]. This DM term is thus rendered inactive, hence there is no spin canting proclivity, and the $z$ component of magnetization therefore remains zero in the entire self consistent calculation. For the starting in-plane AFM order in any direction other than $a$, the $z$ component of magnetization grows, eventually leading to the FM ($z$) order. 

The self consistently determined magnetization and density values are shown in Table V, and the renormalized SOC and orbital moment values in Table VI. The orbital and sublattice averaged magnetization explicitly yields FM ($b$) order with $\langle {\bf m} \rangle_{\rm av}=(0.461,0.461,0)$, which is found to be nearly independent of the octahedral tilting $(t_{m2})$ magnitude. The density values for $yz,xz$ orbitals show strong antiferro orbital ordering. The self consistently determined FM ($b$) order is relevant for the bilayer ruthenate compound. The vanishing $z$ component of magnetic moments is consistent with opposite canting proclivities in the two layers due to opposite sense of $\rm RuO_6$ octahedral tilting of the inter-layer neighboring octahedra. 

\begin{table}[t] \vspace*{-10mm}
\caption{Self consistently determined magnetization and density values for the three orbitals and two sublattices, corresponding to the FM ($b$) order with bare SOC=0.5, $\epsilon_{xy}=-0.2$, $t_{m1}=0.2$, and $t_{m2,m3}=0.15$. The magnetization $z$ component $m_\mu^z=0$ for all three orbitals.} 
\centering 
\begin{tabular}{l c c c} \\  
\hline\hline 
$\mu$ (s) & $m_\mu^x$ & $m_\mu^y$ & $n_\mu$ \\ [0.5ex]
\hline 
$yz$ (A) & 0.654 & 0.506 & 1.169 \\[1ex]
$xz$ (A) & 0.435 & 0.281 & 1.472 \\[1ex]
$xy$ (A) & 0.514 & 0.376 & 1.359 \\[1ex] \hline
\end{tabular} \hspace{10mm}
\begin{tabular}{l c c c} \\  
\hline\hline 
$\mu$ (s) & $m_\mu^x$ & $m_\mu^y$ & $n_\mu$ \\ [0.5ex]
\hline 
$yz$ (B) & 0.281 & 0.435 & 1.472 \\[1ex]
$xz$ (B) & 0.506 & 0.654 & 1.169 \\[1ex]
$xy$ (B) & 0.376 & 0.514 & 1.359 \\[1ex] \hline 
\end{tabular}
\label{table4}
\end{table}

\begin{table}[t]
\caption{Self consistently determined renormalized SOC and orbital magnetic moment values on the two sublattices, corresponding to the FM ($b$) order with same parameters as in Table V.} 
\centering 
\begin{tabular}{l c c c r r c} \\  
\hline\hline 
$s$ & $\lambda_x$ & $\lambda_y$ & $\lambda_z$ & $\langle L_x \rangle$ & $\langle L_y \rangle$ & $\langle L_z \rangle$ \\ [0.5ex]
\hline 
A & 1.204 & 0.585 & 0.744 & $-$0.478 & 0.070 & 0 \\[1ex]
B & 0.585 & 1.204 & 0.744 & 0.070 & $-$0.478 & 0 \\[1ex]
\hline 
\end{tabular}
\label{table5}
\end{table}

\newpage
\section{Conclusions}
Including the orbital off-diagonal spin and charge condensates in the self consistent determination of magnetic order provides a unified framework for understanding the rich magnetic behaviour of the $\rm 4d^4$ compound $\rm Ca_2 Ru O_4$, and illustrates the complex interplay between the different physical elements. These include octahedral flattening (tetragonal distortion) induced dominantly $yz,xz$ moments, SOC induced easy-plane anisotropy, octahedral tilting induced easy-axis anisotropy, SOC induced strong orbital magnetic moments and coupled spin-orbital fluctuations, Coulomb interaction induced strongly anisotropic SOC renormalization, octahedral tilting and rotation induced small canting of moments, and reduced tetragonal distortion induced reorientation transition from AFM to FM order. This transition is weakly tuned by SOC and octahedral tilting, resulting in a tunable magnetic order regime wherein the AFM order is maintained by progressively decreasing octahedral tilting magnitude. The $\rm Ca_2RuO_4$ nanofilm crystal and bilayer ruthenate compound $\rm Ca_3Ru_2O_7$ provide further illustrations of the magnetic order tuning by structural distortions. 

\appendix
\section{SOC induced easy plane anisotropy and magnetic anisotropy energy}
The bare spin-orbit coupling term (for site $i$) can be written in spin space as:
\begin{eqnarray} 
{\cal H}_{\rm SOC} (i) & = & -\lambda {\bf L}.{\bf S} = -\lambda (L_z S_z + L_x S_x + L_y S_y) \nonumber \\ 
&=& \left [ \begin{pmatrix} \psi_{yz \uparrow}^\dagger & \psi_{yz \downarrow}^\dagger \end{pmatrix}
\begin{pmatrix} i \sigma_z \lambda /2 \end{pmatrix} 
\begin{pmatrix} \psi_{xz \uparrow} \\ \psi_{xz \downarrow} \end{pmatrix}
+ \begin{pmatrix} \psi_{xz \uparrow}^\dagger & \psi_{xz \downarrow}^\dagger \end{pmatrix}
\begin{pmatrix} i \sigma_x \lambda /2 \end{pmatrix} 
\begin{pmatrix} \psi_{xy \uparrow} \\ \psi_{xy \downarrow} \end{pmatrix} \right . \nonumber \\
& + & \left . \begin{pmatrix} \psi_{xy \uparrow}^\dagger & \psi_{xy \downarrow}^\dagger \end{pmatrix}
\begin{pmatrix} i \sigma_y \lambda /2 \end{pmatrix} 
\begin{pmatrix} \psi_{yz \uparrow} \\ \psi_{yz \downarrow} \end{pmatrix} \right ] + {\rm H.c.}
\label{soc}
\end{eqnarray}
which explicitly breaks the SU(2) spin rotation symmetry and induces coupled spin-orbital fluctuations. For the orbital angular momentum operators, we have used the matrix representations:
\begin{equation}
L_z = \begin{pmatrix} 0&-i&0 \\ i&0&0 \\ 0&0&0 \end{pmatrix} ,\;\;
L_x = \begin{pmatrix} 0&0&0 \\ 0&0&-i \\ 0&i&0 \end{pmatrix} ,\;\;
L_y = \begin{pmatrix} 0&0&i \\ 0&0&0 \\ -i&0&0 \end{pmatrix} ,\;\;
\end{equation}
in the three-orbital $(yz,xz,xy)$ basis. 

As the orbital ``hopping" terms in Eq. (\ref{soc}) have the same form as spin-dependent hopping terms $i${\boldmath $\sigma . t'_{ij}$}, carrying out the strong-coupling expansion\cite{hc_JMMM_2019} for the $-\lambda L_z S_z$ term to second order in $\lambda$ yields the anisotropic diagonal (AD) intra-site interactions:
\begin{equation}
[H^{(2)}_{\rm eff}]_{\rm AD}^{(z)}(i) = \frac{4 (\lambda/2)^2 }{U} \left [ S_{yz}^z S_{xz}^z - (S_{yz}^x S_{xz}^x  + S_{yz}^y S_{xz}^y) \right ] 
\label{h_eff}
\end{equation}
between moments in the nominally half-filled and magnetically active $yz,xz$ orbitals. Corresponding to an effective single-ion anisotropy (SIA), this term explicitly yields preferential $x-y$ plane ordering for the parallel $yz,xz$ moments enforced by relatively stronger Hund's coupling. The magnetic anisotropy energy (MAE) per site is thus obtained as:
\begin{equation}
\Delta_{\rm MAE} = E_g^{\rm AFM}(\theta=0) - E_g^{\rm AFM}(\theta=\pi/2) = \frac{8(\lambda/2)^2 S^2}{U} = 12.5 \; {\rm meV} 
\end{equation}
where $\theta$ is the polar angle, and we have taken bare SOC value $\lambda = 1.0$, $S=1/2$, $U=8$, and the energy scale unit 200 meV. The MAE involves interplay between SOC and Coulomb interactions, for moderate tetragonal distortion.  

\section{Octahedral tilting and easy-axis anisotropy}
While easy $x-y$ plane anisotropy is directly induced by SOC, interplay between SOC and staggered octahedral tilting in $\rm Ca_2RuO_4$ yields an easy-axis anisotropy along the $\hat{x}+\hat{y}$ direction, which is same as the crystal $b$ direction. Orbital mixing hopping terms between $xy$ and $yz,xz$ orbitals (Eq. \ref{band}) are generated by octahedral tilting. Together with the local SOC spin-flip mixing terms between $xy$ and $yz,xz$ orbitals, these normal NN hopping terms lead to effective spin-dependent NN hopping terms:
\begin{equation}
H_{\rm eff} ' = \sum_{\langle i,j \rangle,\mu} \psi_{i\mu} ^\dagger [-i\makebox{\boldmath $\sigma$}.{\bf t'}] \psi_{j\mu} + {\rm H.c.} 
\label{sdhterms}
\end{equation}
for the magnetically active ($\mu=yz,xz$) orbitals. The hopping terms are bond dependent, with only finite $t_x'$ ($t_y'$) between $xz$ ($yz$) orbital in the $x$ ($y$) direction. Within the usual strong-coupling expansion, the combination of normal ($t$) and spin-dependent ($t_x',t_y'$) hopping terms generates Dzyaloshinski-Moriya (DM) interactions in the effective spin model:
\begin{eqnarray}
[H_{\rm eff}^{(2)}]_{\rm DM}^{(x,y)} &=& \frac{8tt_x'}{U} \sum_{\langle i,j \rangle_x} \hat{x}.({\bf S}_{i,xz} \times {\bf S}_{j,xz}) 
+ \frac{8tt_y'}{U} \sum_{\langle i,j\rangle_y} \hat{y}.({\bf S}_{i,yz} \times {\bf S}_{j,yz}) \nonumber \\ 
& \approx & \frac{8t|t_x'|}{U} \sum_{\langle i,j \rangle} 
(-\hat{x} + \hat{y}).({\bf S}_i \times {\bf S}_j) 
\end{eqnarray}
for $t_x'=-t_y'=-$ive and ${\bf S}_{i,xz} \approx {\bf S}_{i,yz}$ due to the relatively much stronger Hund's coupling. The effective DM axis ($-\hat{x} + \hat{y}$) is along the octahedral tilting axis, which is same as the crystal $a$ axis (Fig. \ref{axes}).  

The above DM interaction term favors spins lying in the plane perpendicular to the DM axis, and  induces spin canting about the DM axis. Intersection of the perpendicular plane $(\phi=\pi/4,z)$ and the SOC-induced easy $x-y$ plane yields $\phi=\pi/4$ as the easy-axis direction. Furthermore, canting of the $yz,xz$ moments about the DM axis yields spin canting in the $z$ direction, as shown in Fig. \ref{canting}(a). 

In close analogy with the above effects of tilting, the staggered octahedral rotation about the crystal $c$ axis leads to orbital mixing hopping terms between $yz,xz$ orbitals on NN sites, and hence to effective spin-dependent NN hopping terms $t_z'$ in Eq. (\ref{sdhterms}). The resulting effective DM term $-(8tt_z'/U)\hat{z}.({\bf S}_i \times {\bf S}_j)$ causes spin canting about the crystal $c$ axis, as shown in Fig. \ref{canting}(b). The easy-axis anisotropy as well as the two spin cantings of the dominant $yz,xz$ moments are confirmed in the full self-consistent calculation (Sec. V).

\section{Orbital off-diagonal condensates in the HF approximation}
The additional contributions in the HF approximation arising from the orbital off-diagonal spin and charge condensates are given below for reference. For the density, Hund's coupling, and pair hopping interaction terms, we obtain (for site $i$):
\begin{eqnarray}
U'' \sum_{\mu < \nu} n_\mu n_\nu & \rightarrow & 
-\frac{U''}{2} \sum_{\mu < \nu} \left [ n_{\mu\nu} \langle n_{\nu\mu} \rangle + \makebox{\boldmath $\sigma$}_{\mu\nu}.\langle \makebox{\boldmath $\sigma$}_{\nu\mu} \rangle \right ] + {\rm H.c.} \nonumber \\
-2J_{\rm H} \sum_{\mu < \nu} {\bf S}_\mu . {\bf S}_\nu 
& \rightarrow & \frac{J_{\rm H}}{4} \sum_{\mu < \nu} 
\left [3\, n_{\mu\nu} \langle n_{\nu\mu} \rangle - \makebox{\boldmath $\sigma$}_{\mu\nu}.\langle \makebox{\boldmath $\sigma$}_{\nu\mu} \rangle \right ] + {\rm H.c.} \nonumber \\
J_{\rm P} \sum_{\mu \ne \nu} a_{\mu\uparrow}^{\dagger} a_{\mu\downarrow}^{\dagger} a_{\nu\downarrow} a_{\nu\uparrow} & \rightarrow &  
\frac{J_{\rm P}}{2} \sum_{\mu < \nu} \left [n_{\mu\nu} \langle n_{\mu\nu} \rangle - \makebox{\boldmath $\sigma$}_{\mu\nu}.\langle \makebox{\boldmath $\sigma$}_{\mu\nu} \rangle \right ] + {\rm H.c.} 
\end{eqnarray}
in terms of the orbital off-diagonal spin ($\makebox{\boldmath $\sigma$}_{\mu\nu}=\psi_\mu^\dagger \makebox{\boldmath $\sigma$} \psi_\nu$) and charge ($n_{\mu\nu} = \psi_\mu^\dagger {\bf 1} \psi_\nu$) operators. The orbital off-diagonal condensates are finite due to the SOC-induced spin-orbital correlations. These additional terms in the HF theory explicitly preserve the SU(2) spin rotation symmetry of the various Coulomb interaction terms. 

Collecting all the spin and charge terms together, we obtain the orbital off-diagonal (OOD) contributions of the Coulomb interaction terms:
\begin{eqnarray}
[{\cal H}_{\rm int}^{\rm HF}]_{\rm OOD} &=& \sum_{\mu < \nu} \left [ \left (-\frac{U''}{2} + \frac{3J_{\rm H}}{4} \right ) 
n_{\mu\nu} \langle n_{\nu\mu} \rangle + \left (\frac{J_{\rm P}}{2}\right ) n_{\mu\nu} \langle n_{\mu\nu} \rangle \right . \nonumber \\
& & - \left . \left (\frac{U''}{2} + \frac{J_{\rm H}}{4} \right )\makebox{\boldmath $\sigma$}_{\mu\nu}.\langle \makebox{\boldmath $\sigma$}_{\nu\mu} \rangle - \left (\frac{J_{\rm P}}{2} \right )\makebox{\boldmath $\sigma$}_{\mu\nu}.\langle \makebox{\boldmath $\sigma$}_{\mu\nu} \rangle \right ] + {\rm H.c.} 
\end{eqnarray}
Separating the condensates $\langle n_{\mu\nu} \rangle = \langle n_{\mu\nu} \rangle^{\rm Re} + i \langle n_{\mu\nu} \rangle^{\rm Im}$ into real and imaginary parts in order to simplify using $\langle n_{\nu\mu}\rangle=\langle n_{\mu\nu}\rangle^*$, and similarly for $\langle \makebox{\boldmath $\sigma$}_{\mu\nu} \rangle$, allows for organizing the OOD charge and spin  contributions into orbital symmetric and anti-symmetric parts:
\begin{eqnarray}
[{\cal H}_{\rm int}^{\rm HF}]_{\rm OOD} &=& -\frac{U''_{\rm c|s}}{2} \sum_{\mu < \nu} \langle n_{\mu\nu} \rangle^{\rm Re} \left [ n_{\mu\nu} + {\rm H.c.} \right ] - \frac{U''_{\rm c|a}}{2} \sum_{\mu < \nu} \langle n_{\mu\nu} \rangle^{\rm Im} \left [ -i n_{\mu\nu} + {\rm H.c.} \right ] \nonumber \\
& - & \frac{U''_{\rm s|s}}{2} \sum_{\mu < \nu} 
\langle \makebox{\boldmath $\sigma$}_{\mu\nu} \rangle^{\rm Re} . 
\left [ \makebox{\boldmath $\sigma$}_{\mu\nu} + {\rm H.c.} \right ] - \frac{U''_{\rm s|a}}{2} \sum_{\mu < \nu} 
\langle \makebox{\boldmath $\sigma$}_{\mu\nu} \rangle^{\rm Im} . 
\left [ -i \makebox{\boldmath $\sigma$}_{\mu\nu} + {\rm H.c.} \right ]
\end{eqnarray}
where the effective interaction terms above are obtained as: 
\begin{eqnarray}
U''_{\rm c|a} &=& U''_{\rm s|a} = U'' - J_{\rm H}/2 = U - 3J_{\rm H} \nonumber \\
U''_{\rm s|s} &=& U'' + 3J_{\rm H}/2 = U - J_{\rm H} \nonumber \\
U''_{\rm c|s} &=& U''-5J_{\rm H}/2 = U-5J_{\rm H} 
\end{eqnarray}
using $J_{\rm P}=J_{\rm H}$. While the effective interaction $U''_{\rm s|s}$ (spin term, symmetric part) is enhanced relative to $U''$, the corresponding charge term interaction $U''_{\rm c|s}$ vanishes for $J_{\rm H}=U/5$.


\begin{thebibliography}{99}

\bibitem{fatuzzo_PRB_2015} C. G. Fatuzzo, M. Dantz, S. Fatale, P. Olalde-Velasco, N. E. Shaik, B. Dalla Piazza, S. Toth, J. Pelliciari, R. Fittipaldi, A. Vecchione, N. Kikugawa, J. S. Brooks, H. M. R\o{}nnow, M. Grioni, Ch. R\"uegg, T. Schmitt, and J. Chang, \textit{Spin-Orbit-Induced
Orbital Excitations in $\rm Sr_2RuO_4$ and $\rm Ca_2RuO_4$: A Resonant Inelastic X-ray Scattering Study}, Phys. Rev. B {\bf 91}, 155104 (2015).

\bibitem{zhang_PRL_2016} G. Zhang, E. Gorelov, E. Sarvestani, and E. Pavarini, \textit{Fermi Surface of $\rm Sr_2RuO_4$: Spin-Orbit and Anisotropic Coulomb Interaction Effects}, Phys. Rev. Lett. {\bf 116}, 106402 (2016).

\bibitem{jain_NATPHY_2017} A. Jain, M. Krautloher, J. Porras, G. H. Ryu, D. P. Chen, D. L. Abernathy, J. T. Park, A. Ivanov, J. Chaloupka, G. Khaliullin, B. Keimer, and B. J. Kim, \textit{Higgs Mode and Its Decay in a Two-Dimensional Antiferromagnet}, Nat. Phys. {\bf 13}, 633 (2017).

\bibitem{souliou_PRL_2017} S.-M. Souliou, J. Chaloupka, G. Khaliullin, G. Ryu, A. Jain, B. J. Kim, M. Le Tacon, and B. Keimer, \textit{Raman Scattering from Higgs Mode Oscillations in the Two-Dimensional Antiferromagnet $\rm Ca_2RuO_4$}, Phys. Rev. Lett. {\bf 119}, 067201 (2017).

\bibitem{zhang_PRB_2017} G. Zhang and E. Pavarini, \textit{Mott transition, Spin-Orbit Effects, and Magnetism in $\rm Ca_2RuO_4$}, Phys. Rev. B {\bf 95}, 075145 (2017).

\bibitem{porter_PRB_2018} D. G. Porter, V. Granata, F. Forte, S. Di Matteo, M. Cuoco, R. Fittipaldi, A. Vecchione, and A. Bombardi, \textit{Magnetic Anisotropy and Orbital Ordering in $\rm Ca_2 RuO_4$}, Phys. Rev. B {\bf 98}, 125142 (2018).

\bibitem{das_PRX_2018} L. Das, F. Forte, R. Fittipaldi, C. G. Fatuzzo, V. Granata, O. Ivashko, M. Horio, F. Schindler, M. Dantz, Yi Tseng, D. E. McNally, H. M. R\o{}nnow, W. Wan, N. B. Christensen, J. Pelliciari, P. Olalde-Velasco, N. Kikugawa, T. Neupert, A. Vecchione, T. Schmitt, M. Cuoco, and J. Chang, \textit{Spin-Orbital Excitations in $\rm Ca_2RuO_4$
Revealed by Resonant Inelastic X-Ray Scattering}, Phys. Rev. X {\bf 8}, 011048 (2018).

\bibitem{dietl_APL_2018} C. Dietl, S. K. Sinha, G. Christiani, Y. Khaydukov, T. Keller, D. Putzky, S. Ibrahimkutty, P. Wochner, G. Logvenov, P. A. van Aken, B. J. Kim, and B. Keimer, \textit{Tailoring the Electronic Properties of $\rm Ca_2RuO_4$ via Epitaxial Strain}, Appl. Phys. Lett. {\bf 112}, 031902 (2018).

\bibitem{kim_PRL_2018} M. Kim, J. Mravlje, M. Ferrero, O. Parcollet, and A. Georges, \textit{Spin-Orbit Coupling and Electronic Correlations in $\rm Sr_2RuO_4$}, Phys. Rev. Lett. {\bf 120}, 126401 (2018).

\bibitem{feldmaier_arxiv_2019} T. Feldmaier, P. Strobel, M. Schmid, P. Hansmann, and M. Daghofer, \textit{Excitonic Magnetism at the Intersection
of Spin-Orbit Coupling and Crystal-Field Splitting}, arXiv:1910.13977 (2019).

\bibitem{gretarsson_PRB_2019} H. Gretarsson, H. Suzuki, H. Kim, K. Ueda, M. Krautloher, B. J. Kim, H. Yavaş, G. Khaliullin, and B. Keimer, \textit{Observation of Spin-Orbit Excitations and Hund's multiplets in $\rm Ca_2RuO_4$}, Phys. Rev. B {\bf 100}, 045123 
(2019).

\bibitem{zhang_PRB_2020} G. Zhang and E. Pavarini, \textit{Higgs Mode and Stability of $xy$-Orbital Ordering in $\rm Ca_2RuO_4$}, Phys. Rev. B {\bf 101}, 205128 (2020).

\bibitem{arx_arxiv_2020} K. von Arx, F. Forte, M. Horio, V. Granata, Q. Wang, L. Das, Y. Sassa, R. Fittipaldi, C. G. Fatuzzo, O. Ivashko, Y. Tseng, E. Paris, A. Vecchione, T. Schmitt, M. Cuoco, J. Chang, \textit{Comparative Resonant Inelastic X-ray Scattering Study of $\rm Ca_2RuO_4$ and $\rm Ca_3Ru_2O_7$}, arXiv:2004.13391 (2020).

\bibitem{nobukane_SREP_2020} H. Nobukane, K.  Yanagihara, Y. Kunisada, Y. Ogasawara, K. Isono, K. Nomura, K. Tanahashi, T. Nomura, T. Akiyama, and S. Tanda, \textit{Co-appearance of Superconductivity and Ferromagnetism in a $\rm Ca_2RuO_4$ Nanofilm Crystal}, Sci. Rep. {\bf 10}, 3462 (2020).

\bibitem{nakatsuji_PRL_2000} S. Nakatsuji and Y. Maeno, \textit{Quasi-Two-Dimensional Mott Transition System $\rm Ca_{2-x} Sr_x RuO_4$}, Phys. Rev. Lett. {\bf 84}, 2666 (2000).

\bibitem{friedt_PRB_2001} O. Friedt, M. Braden, G. Andr{\'e}, P. Adelmann, S. Nakatsuji, and Y. Maeno, \textit{Structural and Magnetic Aspects of the Metal-Insulator Transition in
$\rm Ca_{2-x} Sr_x RuO_4$}, Phys. Rev. B {\bf 63}, 174432 (2001).

\bibitem{fang_PRB_2001} Z. Fang and K. Terakura. \textit{Magnetic Phase Diagram of $\rm Ca_{2-x} Sr_x RuO_4$ Governed by Structural Distortions}, Phys. Rev. B {\bf 64}, 020509 (2001).

\bibitem{fang_PRB_2004} Z. Fang, N. Nagaosa, and K. Terakura, \textit{Orbital-Dependent Phase Control in $\rm Ca_{2-x} Sr_x RuO_4$} ($0 \lesssim$ x $\lesssim 0.5$), Phys. Rev. B {\bf 69}, 045116 (2004).

\bibitem{peng_PRB_2010} J. Peng, Z. Qu, B. Qian, D. Fobes, T. Liu, X. Wu, H. M. Pham, L. Spinu, and Z. Q. Mao, \textit{Interplay Between the Lattice and Spin Degrees of Freedom in $\rm (Sr_{1-x}Ca_x)_3 Ru_2 O_{7}$}, Phys. Rev. B {\bf 82}, 024417 (2010).

\bibitem{nakatsuji_JPSP_1997} S. Nakatsuji, S. ichi Ikeda, and Y. Maeno, \textit{$\rm Ca_2RuO_4$: New Mott Insulators of Layered Ruthenate}, J. Phys. Soc. Jpn {\bf 66}(7), 1868 (1997).

\bibitem{braden_PRB_1998} M. Braden, G. Andr{\'e}, S. Nakatsuji, and Y. Maeno, \textit{Crystal and Magnetic Structure of $\rm Ca_2RuO_4$: Magnetoelastic Coupling and the Metal-Insulator Transition}, Phys. Rev. B {\bf 58}, 847 (1998).

\bibitem{alexander_PRB_1999} C. S. Alexander, G. Cao, V. Dobrosavljevic, S. McCall, J. E. Crow, E. Lochner, and R. P. Guertin, \textit{Destruction of the Mott Insulating Ground State of $\rm Ca_2RuO_4$ by a Structural Transition}, Phys. Rev. B {\bf 60}, R8422 (1999).

\bibitem{gorelov_PRL_2010} E. Gorelov, M. Karolak, T. O. Wehling, F. Lechermann, A. I. Lichtenstein, and E. Pavarini, \textit{Nature of the Mott Transition in $\rm Ca_2RuO_4$}, Phys. Rev. Lett. {\bf 104}, 226401 (2010).

\bibitem{Kunkemoller_PRL_2015} S. Kunkem\"oller, D. Khomskii, P. Steffens, A. Piovano, A. A. Nugroho, and M. Braden, \textit{Highly Anisotropic Magnon Dispersion in $\rm Ca_2 Ru O_4$: Evidence for Strong Spin Orbit Coupling}, Phys. Rev. Lett. {\bf 115}, 247201 (2015).

\bibitem{nakatsuji_PRB_2000} S. Nakatsuji and Y. Maeno, \textit{Switching of Magnetic Coupling by a Structural Symmetry Change Near the Mott Transition in $\rm Ca_{2-x} Sr_x RuO_4$}, Phys. Rev. B {\bf 62}, 6458 (2000).

\bibitem{nakamura_PRB_2002} F. Nakamura, T. Goko, M. Ito, T. Fujita, S. Nakatsuji, H. Fukazawa, Y. Maeno, P. Alireza, D. Forsythe, and S. R. Julian, \textit{From Mott insulator to ferromagnetic metal: A pressure study of $\rm Ca_2 RuO_4$}, Phys. Rev. B {\bf 65}, 220402(R) (2002).

\bibitem{steffens_PRB_2005} P. Steffens, O. Friedt, P. Alireza, W. G. Marshall, W. Schmidt, F. Nakamura, S. Nakatsuji, Y. Maeno, R. Lengsdorf, M. M. Abd-Elmeguid, and M. Braden, \textit{High-Pressure Diffraction Studies on $\rm Ca_2 RuO_4$}, Phys. Rev. B {\bf 72},
094104 (2005).

\bibitem{steffens_PRB_2011} P. Steffens, O. Friedt, Y. Sidis, P. Link, J. Kulda, K. Schmalzl, S. Nakatsuji, M. Braden, \textit{Magnetic Excitations in the Metallic Single-Layer Ruthenates
$\rm Ca_{2-x} Sr_x RuO_4$ Studied by Inelastic Neutron Scattering}, Phys. Rev. B {\bf 83}, 054429 (2011).

\bibitem{nakamura_SREP_2013} F. Nakamura, M. Sakaki, Y. Yamanaka, S. Tamaru, T. Suzuki, and Y. Maeno, \textit{Electric-Field-Induced Metal Maintained by Current of the Mott Insulator $\rm Ca_2 Ru O_4$}, Sci. Rep. {\bf 3}, 2536 (2013).

\bibitem{okazaki2_JPSJ_2013} R. Okazaki, Y. Nishina, Y. Yasui, F. Nakamura, T. Suzuki, and I. Terasaki, \textit{Current-Induced Gap Suppression in the Mott Insulator $\rm Ca_2 Ru O_4$}, J. Phys. Soc. Jpn. {\bf 82}, 103702 (2013).

\bibitem{liebsch_PRL_2007} A. Liebsch and H. Ishida, \textit{Subband Filling and Mott Transition in $\rm Ca_{2-x} Sr_x RuO_4$}, Phys. Rev. Lett. {\bf 98}, 216403 (2007).

\bibitem{khaliullin_PRL_2013} G. Khaliullin, \textit {Excitonic Magnetism in Van Vleck--type ${d}^{4}$ Mott Insulators}, Phys. Rev. Lett. {\bf 111}, 197201 (2013).

\bibitem{akbari_PRB_2014} A. Akbari and G. Khaliullin, \textit{Magnetic Excitations in a Spin-Orbit-Coupled ${d}^{4}$ Mott Insulator on the Square Lattice}, Phys. Rev. B {\bf 90}, 035137 (2014).

\bibitem{sutter_NATCOM_2017} D. Sutter, C. G. Fatuzzo, S. Moser, M. Kim, R. Fittipaldi, A. Vecchione, V. Granata, Y. Sassa, F. Cossalter, G. Gatti, M. Grioni, H. M. R{\o}nnow, N. C. Plumb, C. E. Matt, M. Shi, M. Hoesch, T. K. Kim, T.-R. Chang, H.-T. Jeng, C. Jozwiak, A. Bostwick, E. Rotenberg, A. Georges, T. Neupert, and J. Chang, \textit{Hallmarks of Hund's Coupling in the Mott Insulator $\rm Ca_2 Ru O_4$}, Nat. Comm. {\bf 8}, 15176 (2017).

\bibitem{ruthenate_one_2020} S. Mohapatra and A. Singh, \textit{Magnetic Reorientation Transition in a Three Orbital Model for $\rm Ca_2 Ru O_4$ -- Interplay of Spin-Orbit Coupling, Tetragonal Distortion, and Coulomb Interactions}, arXiv:2006.02114 (2020).

\bibitem{mohapatra_PRB_2019} S. Mohapatra, S. Aditya, R. Mukherjee, and A. Singh, \textit{Octahedral Tilting Induced Isospin Reorientation Transition in Iridate Heterostructures}, Phys. Rev. B {\bf 100}, 140409(R) (2019).

\bibitem{yoshida_PRB_2005} Y. Yoshida, S.-I. Ikeda, H. Matsuhata, N. Shirakawa, C. H. Lee, and S. Katano, \textit{Crystal and Magnetic Structure of $\rm Ca_3Ru_2O_7$}, Phys. Rev. B {\bf 72}, 054412 (2005).  

\bibitem{cao_PRB_2003} G. Cao, L. Balicas, Y. Xin, J. E. Crow, and C. S. Nelson, \textit{Quantum Oscillations, Colossal Magnetoresistance, and the Magnetoelastic Interaction in Bilayered $\rm Ca_3Ru_2O_7$}, Phys. Rev. B {\bf 67}, 184405 (2003).

\bibitem{yoshida_PRB_2004} Y. Yoshida, I. Nagai, S. I. Ikeda, N. Shirakawa, M. Kosaka, and N. Mori, \textit{Quasi-Two-Dimensional Metallic Ground State of $\rm Ca_3Ru_2O_7$}, Phys. Rev. B {\bf 69}, 220411(R) (2004).

\bibitem{mccall_PRB_2003} S. McCall, G. Cao, and J. E. Crow, \textit{Impact of Magnetic Fields on Anisotropy in $\rm Ca_3Ru_2O_7$}, Phys. Rev. B {\bf 67}, 094427 (2003).

\bibitem{lin_PRL_2005} X. N. Lin, Z. X. Zhou, V. Durairaj, P. Schlottmann, and G. Cao, \textit{Colossal Magnetoresistance by Avoiding a Ferromagnetic State in the Mott System $\rm Ca_3Ru_2O_7$}, Phys. Rev. Lett. {\bf 95}, 017203 (2005).

\bibitem{bao_PRL_2008} W. Bao, Z. Q. Mao, Z. Qu, and J. W. Lynn, \textit{Spin Valve Effect and Magnetoresistivity in Single Crystalline $\rm Ca_3Ru_2O_7$}, Phys. Rev. Lett. {\bf 100}, 247203 (2008).

\bibitem{zhu_PRL_2016} M. Zhu, J. Peng, T. Zou, K. Prokes, S. D. Mahanti, T. Hong, Z. Q. Mao, G. Q. Liu, and X. Ke, \textit{Colossal Magnetoresistance in a Mott Insulator via Magnetic Field-Driven Insulator-Metal Transition}, Phys. Rev. Lett. {\bf 116}, 216401 (2016).

\bibitem{hc_JMMM_2019} S. Mohapatra and A. Singh, \textit{Spin Waves and Stability of Zigzag Order in the Hubbard Model with Spin-Dependent Hopping Terms: Application to the Honeycomb Lattice Compounds $\rm Na_2IrO_3$ and $\alpha$-$\rm RuCl_3$}, J. Magn. Magn. Mater {\bf 479}, 229 (2019).

\end{thebibliography}
\end{document}